# On the energetics of a tidally oscillating convective flow

Caroline Terquem [1,2,3]★

[1] *Rudolf Peierls Centre for Theoretical Physics, University of Oxford, Parks Road, Oxford OX1 3PU, UK*
[2] *University College, Oxford OX1 4BH, UK*
[3] *Institut d'Astrophysique de Paris, Sorbonne Université, CNRS, UMR 7095, 98 bis boulevard Arago, F-75014 Paris, France*

**ABSTRACT**

This paper examines the energetics of a convective flow subject to an oscillation with a period $t_{osc}$ much smaller than the convective time-scale $t_{conv}$, allowing for compressibility and uniform rotation. We show that the energy of the oscillation is exchanged with the kinetic energy of the convective flow at a rate $D_R$ that couples the Reynolds stress of the oscillation with the convective velocity gradient. For the equilibrium tide and inertial waves, this is the only energy exchange term, whereas for $p$ modes there are also exchanges with the potential and internal energy of the convective flow. Locally, $|D_R| \sim u'^2/t_{conv}$, where $u'$ is the oscillating velocity. If $t_{conv} \ll t_{osc}$ and assuming mixing length theory, $|D_R|$ is $(\lambda_{conv}/\lambda_{osc})^2$ smaller, where $\lambda_{conv}$ and $\lambda_{osc}$ are the characteristic scales of convection and the oscillation. Assuming local dissipation, we show that the equilibrium tide lags behind the tidal potential by a phase $\delta(r) \sim r\omega_{osc}/(g(r)t_{conv}(r))$, where $g$ is the gravitational acceleration. The equilibrium tide can be described locally as a harmonic oscillator with natural frequency $(g/r)^{1/2}$ and subject to a damping force $-u'/t_{conv}$. Although $\delta(r)$ varies by orders of magnitude through the flow, it is possible to define an average phase shift $\bar{\delta}$ which is in good agreement with observations for Jupiter and some of the moons of Saturn. Finally, $1/\bar{\delta}$ is shown to be equal to the standard tidal dissipation factor.

**Key words:** convection – hydrodynamics – Sun: general – planets and satellites: dynamical evolution and stability – planet–star interactions – binaries: close.

## 1 INTRODUCTION

The circularization of stellar binaries and the orbital evolution of the moons of giant planets give very good constraints on the amount of energy dissipation in bodies in which tides are excited. Both solar-type stars and giant planets have extensive convective envelopes. Starting with Zahn (1966), it has commonly been assumed that convection acts as a turbulent viscosity which damps tidal oscillations. In this picture, the rate at which kinetic energy per unit mass is exchanged between tidal oscillations and the convective flow is $D_R = \langle u_i' u_j' \rangle (\partial V_i/\partial x_j)$, where $\boldsymbol{u}'$ is the convective velocity and $\boldsymbol{V}$ is that of the tides, and mixing length theory is used to express $\langle u_i' u_j' \rangle$ in terms of a turbulent viscosity and the shear associated with the tidal velocity. This *assumes* that $D_R$ has the (negative) sign required for energy to be transferred from tidal oscillations to the convective flow, leading to dissipation of the tides. This requires the energy of the background shear flow (tides) to be fed into convective motions. This is by no means a trivial assumption, and there are counter examples of convection acting as a negative viscosity (Starr 1968). It is also well known that, in the Sun, energy is extracted from convection to be fed into the shear associated with differential rotation. However, Zahn's theory is very successful at reproducing the circularization time-scales inferred from observations in cases where the convective turnover time-scale $t_{conv}$ is small compared to the period of the oscillations $t_{osc}$ (Verbunt & Phinney 1995).

In the opposite regime $t_{conv} \gg t_{osc}$, it has been argued that the turbulent viscosity should be reduced because convective eddies cannot exchange momentum with their environment during a tidal period (Zahn 1966; Goldreich & Nicholson 1977). This results in rates of energy dissipation orders of magnitude too small to account for observations. The discrepancy between theory and observations is particularly severe for Jupiter, where $t_{conv}/t_{osc}$ is larger than $10^3$ in the whole of the envelope when considering the interaction with Io.

However, the very fact that there is hardly any convective transport during a tidal period when $t_{conv} \gg t_{osc}$ actually invalidates the model of convection acting as a turbulent viscosity (Terquem 2021). In this regime, the rate at which kinetic energy is exchanged between tidal oscillations and the convective flow is still $D_R$ as written above, but with $\boldsymbol{u}'$ being the velocity of the tides and $\boldsymbol{V}$ that of the convective flow: the role of the fluctuations and that of the mean flow are reversed. This is because the Reynolds stress $-\rho \langle u_i' u_j' \rangle$ is always associated with the fastest varying component of the flow. Here again, assuming that $D_R$ has the (positive) sign required for energy to be transferred from the tides to convection, Terquem & Martin (2021) obtained rates of energy dissipation in agreement with observations for the circularization of solar-type stars.

★ E-mail: caroline.terquem@physics.ox.ac.uk

The fact that observations can be accounted for by assuming that energy is systematically transferred from the tides to convection, both when the tides act as the fluctuations or as the mean flow, is compelling. However, it is worth noting that in either case there is no proof so far that tidal energy is always dissipated.

The analysis in Terquem (2021) was restricted to incompressible and non-rotating flows. The present paper aims at generalizing this analysis by including compressibility and uniform rotation. The addition of compressibility will enable the application of the formalism to the case where the oscillations are pressure modes, which are associated with time-scales much shorter than the convective time-scales in stars. It is believed that they are damped through their interaction with the convective flow, and this is usually modelled using mixing length theory. However, this has been found not to be in good agreement with numerical simulations (Basu 2016), which is not surprising in the context of the new formalism which applies when $t_{osc} \ll t_{conv}$. The addition of uniform rotation will also enable the application of the formalism to inertial waves. This is important because, if the tidal frequency is smaller than twice the rotation frequency, the response of a rotating star or giant planet to a tidal perturbation is the superimposition of an equilibrium tide and propagating inertial waves. It has been assumed in previous studies that these waves are damped through their interaction with a turbulent convective viscosity. However, like for the equilibrium tide, damping of inertial waves cannot be described by mixing length theory when $t_{osc} \ll t_{conv}$.

This paper examines in full generality in which form energy is transferred between an oscillation and convection. We take into account all the energy stores and examine all the routes that the energy of the oscillation could potentially travel through.

We then focus on tidal oscillations and, assuming local dissipation, calculate the phase by which the tide lags behind the perturbing potential. We also explore the damped harmonic oscillator model for the equilibrium tide.

The plan of the paper is as follows. In Section 2, we write the energy conservation equations for the flow as a whole, without separating the oscillation from the convective flow. Assuming $t_{osc} \ll t_{conv}$, we then perform a Reynolds decomposition and write the kinetic, potential, and internal energy conservation equations for the mean convective flow and for the oscillation separately in Section 3. These equations are averaged first over the time-scale $t_{osc}$ and then over a time-scale long compared to $t_{conv}$. We show that the energy of the oscillation is exchanged with the kinetic energy of the convective flow at a rate $D_R$ per unit mass that couples the Reynolds stress of the oscillation with the convective velocity gradient. If the oscillation is the equilibrium tide or an inertial wave, $D_R$ is the only term that exchanges energy between the oscillation and the convective flow, and it is always balanced by the work done by the tidal force, whether this is positive or negative. If the oscillation is a *p* mode, there is an additional exchange with the potential and internal energy of the convective flow, because of compressibility. In Section 4, we discuss how to express $D_R$ in terms of the flow velocities. We review the standard cases of viscous and turbulent shear flows, and examine oscillations with $t_{osc}$ either small or large compared to $t_{conv}$. This makes it clear that many of the questions that arise about the direction of energy transfer when the tides are fast are also relevant when the tides are slow and play the role of the mean flow. The two cases should therefore be examined together. When $t_{conv} \gg t_{osc}$, $|D_R| \sim u'^2/t_{conv}$, where $u'$ is the oscillating velocity. If $t_{conv} \ll t_{osc}$ and assuming mixing length theory, $|D_R|$ is $(\lambda_{conv}/\lambda_{osc})^2$ smaller, where $\lambda_{conv}$ is the mixing length and $\lambda_{osc}$ is the spatial scale on which the oscillation varies. This applies to *p* modes, the equilibrium tide and inertial waves. In Section 5, we focus on tidal oscillations and assume local dissipation. We show that the phase shift between the oscillation and the tidal potential varies by orders of magnitude through the flow. For the equilibrium tide, which we model as a harmonic oscillator, it is however possible to define a mean phase shift for the flow which can be compared to the values derived from observations for Jupiter and Saturn. The inverse of this mean phase shift is equal to the standard tidal dissipation factor $Q = 2\pi E'_p/\Delta E$, where $E'_p$ is the potential energy in the tide and $\Delta E$ is the total energy dissipated during a tidal period. Finally, we summarize and discuss our results in Section 6.

## 2 THE DIFFERENT FORMS OF ENERGY

We consider a body (solar-type star or giant planet) in uniform rotation which has an envelope in which energy is transported by convection. We note $\boldsymbol{\Omega}$ the angular velocity vector of the body, $\boldsymbol{u}$ the velocity of the gas in the rotating frame, $\rho$ its density and $p$ its pressure. The body is subject to a tidal force per unit mass $\boldsymbol{f}_t = -\nabla \Psi_t$, where $\Psi_t$ is the tidal potential in the rotating frame.

### 2.1 Conservation of energy

To help the discussion, we first recall the equations expressing conservation of energy for the flow in the convective envelope without separating tidal oscillations and convective motions. The flow satisfies the mass conservation equation

$$\frac{\partial \rho}{\partial t} + \frac{\partial}{\partial x_j}\left(\rho u_j\right) = 0, \tag{1}$$

and Navier–Stokes equation, which *i*-component in Cartesian coordinates is

$$\rho \frac{\partial u_i}{\partial t} + \rho u_j \frac{\partial u_i}{\partial x_j} = -\frac{\partial p}{\partial x_i} + \rho g_i + \rho f_{c,i} + \rho f_{t,i} + \frac{\partial \sigma_{ij}}{\partial x_j}, \tag{2}$$

where $g_i$ is the (negative) acceleration due to the gravity of the body itself in the *i*-direction, $\boldsymbol{f}_c = -2\boldsymbol{\Omega} \times \boldsymbol{u}$ is the Coriolis force per unit mass, $\sigma_{ij}$ is the viscous stress tensor and repeated indices are summed over. In the cases of interest, the centrifugal force is very small compared to the gravitational force, into which it can be subsumed. For example, in the Sun, the centrifugal force is about $10^5$ times smaller than the gravitational force (Miesch 2005). For this reason, it is neglected here. In a solar-type star, almost all of the mass is in the radiative core, so that $g \simeq -GM/r^2$, where $M$ is the total mass of the star. In principle, to calculate $g$ in the envelope of a giant planet, self-gravity has to be taken into account. However, in the parts of the envelope where tides have a significant amplitude, $g$ is also well approximated by $-GM/r^2$, where $M$ is the total mass of the planet. This approximation will therefore be used thereafter. We note $\Psi_0$ the associated gravitational potential, defined through $\boldsymbol{g} = -\nabla \Psi_0$ (i.e. in spherical coordinates, $\Psi_0 = -GM/r$).

Multiplying equation (2) by $u_i$, summing up over $i$ and using equation (1) yields the equation for kinetic energy conservation

$$\frac{\partial}{\partial t}\left(\frac{1}{2}\rho u^2\right) = -\frac{\partial}{\partial x_j}\left[\left(\frac{1}{2}\rho u^2 + p\right)u_j - \sigma_{ij}u_i\right] + p\nabla \cdot \boldsymbol{u} + \rho \boldsymbol{g} \cdot \boldsymbol{u} + \rho \boldsymbol{f}_t \cdot \boldsymbol{u} - \sigma_{ij}\frac{\partial u_i}{\partial x_j}. \tag{3}$$

An equation for the gravitational potential energy is obtained by multiplying equation (1) by the potential energy per unit mass $\Psi = \Psi_0 + \Psi_t$. Assuming $\Psi_0$ to be independent of time, this yields

$$\frac{\partial}{\partial t}(\rho \Psi) = -\frac{\partial}{\partial x_j}(\rho \Psi u_j) - \rho \boldsymbol{g} \cdot \boldsymbol{u} - \rho \boldsymbol{f}_t \cdot \boldsymbol{u} + \rho \frac{\partial \Psi_t}{\partial t}. \tag{4}$$

Equations (3) and (4) show that potential and kinetic energies are exchanged through the work $\rho(\boldsymbol{g} + \boldsymbol{f}_t) \cdot \boldsymbol{u}$ done by the total gravitational force, which includes both the tidal force and the self-gravitational force. The term $\rho(\partial \Psi_t/\partial t)$ represents the excess or deficit of potential energy due to the time dependence of the tidal potential. Although it does not appear to involve work done, we will see in Section 3.7 that, when integrated over the envelope, its average over a tidal period is actually equal to the average work done by the tidal force.

An equation for the internal energy is obtained from the first law of thermodynamics

$$\frac{\partial e_{\text{int}}}{\partial t} = -\frac{\partial}{\partial x_j}(e_{\text{int}} u_j) - p\nabla \cdot \boldsymbol{u} + \sigma_{ij}\frac{\partial u_i}{\partial x_j} - \nabla \cdot \boldsymbol{q}, \tag{5}$$

where $e_{\text{int}}$ is the internal energy per unit volume and $\boldsymbol{q}$ is the radiative flux of thermal energy. Convective and radiative transport of energy are contained in the first and last terms on the right-hand side, respectively. Equations (3) and (5) show that kinetic and internal energies are exchanged through viscous dissipation and through the work $p\nabla \cdot \boldsymbol{u}$ done by the pressure force.

By adding equations (3), (4), and (5), we obtain a conservation equation for the total energy per unit volume $e_{\text{tot}} = \rho u^2/2 + \rho \Psi + e_{\text{int}}$

$$\frac{\partial e_{\text{tot}}}{\partial t} = -\frac{\partial}{\partial x_j}\left[(e_{\text{tot}} + p)u_j - \sigma_{ij}u_i\right] + \rho \frac{\partial \Psi_t}{\partial t} - \nabla \cdot \boldsymbol{q}. \tag{6}$$

The work done by the tidal force does not appear as a source term for the total mechanical energy, no more than the work done by the buoyancy force does, because the tidal potential is included in the potential energy.

## 2.2 Energy dissipation

We now integrate equation (6) over the whole volume of the convective envelope. The divergence term on the right-hand side becomes an integral over the surfaces. The contribution from the outer surface is negligible as $\rho$ and $p$ are very small there. At the inner surface, the convective velocity vanishes, the velocity of the equilibrium tide is very small, and the radial velocity of inertial waves has to vanish to satisfy the boundary condition with either a solid core or a radiative layer. The contribution from the inner surface is therefore also negligible, and the surface integral can be ignored.

In the absence of tides, transport of energy throughout the envelope adjusts itself so that a steady state is maintained (i.e. $e_{\text{tot}}$ integrated over the envelope is constant) over time-scales long compared to that of the flow and small compared to the time-scale on which the internal structure of the body evolves. When tides are present, the total energy has contribution from the tidal oscillation. Averaged over a tidal period, this contribution is constant over a time-scale short compared to the tidal evolution time-scale of the binary. Therefore, over such time-scales, the integral of the term on the left-hand side of equation (6) is zero and we then obtain

$$\int_{\mathcal{V}} \left\langle \rho \frac{\partial \Psi_t}{\partial t} \right\rangle dv + E_{\text{core}} = \int_{\mathcal{S}} \boldsymbol{F}_{\text{rad}} \cdot d\boldsymbol{s}, \tag{7}$$

where $\mathcal{V}$ is the volume of the convective envelope, $\boldsymbol{F}_{\text{rad}}$ is the radiative flux through the outer surface $\mathcal{S}$, and $E_{\text{core}}$ is the thermal energy entering through the inner surface (due to nuclear energy production in a star and other mechanisms in a giant planet). The brackets denote an average over the tidal period.

The equation above shows that, if the tidal potential results in an excess of potential energy (i.e. the integral on the left-hand side is positive), it has to be ultimately radiated away along with the energy $E_{\text{core}}$ produced in the core of the body. As potential energy cannot be converted directly into internal energy, which is the only form of energy which can be radiated away, it first has to be converted into kinetic energy.

Equation (7) can also be written as

$$\int_{\mathcal{V}} \left\langle \rho \frac{\partial \Psi_t}{\partial t} \right\rangle dv = \delta \left( \int_{\mathcal{S}} \boldsymbol{F}_{\text{rad}} \cdot d\boldsymbol{s} \right), \tag{8}$$

where the term on the right-hand side is the perturbation to the net surface flux due to the tidal oscillation, which has contribution from the perturbation to the flux itself, but also from the perturbation to the surface normal and to the surface area. An explicit calculation shows that this term is proportional to the Lagrangian variation of the radial flux at the surface, which itself depends on the radial component of the tidal displacement there (Dziembowski 1977; Bunting & Terquem 2021). Numerical simulations aiming at calculating tidal dissipation should therefore allow the outer surface of the flow to oscillate under tidal forcing. Enforcing rigid boundaries may prevent the energy to be released and create spurious results.

## 3 EQUATIONS FOR THE OSCILLATION AND THE MEAN FLOW

The response of a rotating convective flow to tidal forcing includes a non-wavelike part (equilibrium tide) to which are superimposed propagating inertial waves when the tidal frequency is smaller than twice the rotation frequency. In addition to tidally forced oscillations, a convective flow can also support propagating pressure waves.

In the analysis presented below, we consider a forced oscillation in a compressible convective flow in uniform rotation. The oscillation can therefore be identified with either an equilibrium tide, the sum of an equilibrium tide and a propagating inertial wave if the rotation frequency is large enough, or a propagating pressure wave (for which $f_t$ has to be thought of as the force that excites the wave rather than a tidal force).

We note $t_{\rm osc}$ and $\omega_{\rm osc}$ the period and frequency, respectively, of the oscillation in the rotating frame. For tidal oscillations, we have $\omega_{\rm osc} = 2|\omega_{\rm orb} - \Omega|$, where $\omega_{\rm orb}$ is the orbital frequency of the binary (we only consider circular orbits). In this section, we assume $t_{\rm osc} \ll t_{\rm conv}$, where $t_{\rm conv}$ is the characteristic time-scale associated with convective motions. We will however also discuss the case where the oscillation time-scale is the longest one in the following section.

### 3.1 Reynolds decomposition

We use the Reynolds decomposition in which the total velocity in the rotating frame is written as $\boldsymbol{u}(\boldsymbol{r},t) = \boldsymbol{V}(\boldsymbol{r},t) + \boldsymbol{u}'(\boldsymbol{r},t)$, where $\boldsymbol{u}'$ is the velocity associated with the periodic oscillation in the rotating frame. In other words, if we note $\langle \ldots \rangle$ a time-average over $t_{\rm osc}$, then $\langle \boldsymbol{u}'(\boldsymbol{r},t) \rangle = \boldsymbol{0}$. This defines $\boldsymbol{V}(\boldsymbol{r},t) \equiv \langle \boldsymbol{u}(\boldsymbol{r},t) \rangle$ as the mean velocity, and this is the convective velocity in the rotating frame. As evidenced by the Sun, convection in the presence of rotation may induce differential rotation. This is not always the case, however, as Jupiter for example is mostly in rigid body rotation. When differential rotation is present, then it can be thought of as being included in $\boldsymbol{V}$ (e.g. Durney & Spruit 1979).

A Reynolds decomposition can also be made for the pressure $p(\boldsymbol{r},t) = p_0(\boldsymbol{r}) + \delta p(\boldsymbol{r},t) + p'(\boldsymbol{r},t)$ and the mass density $\rho(\boldsymbol{r},t) = \rho_0(\boldsymbol{r}) + \delta\rho(\boldsymbol{r},t) + \rho'(\boldsymbol{r},t)$, with $\langle p'(\boldsymbol{r},t) \rangle = \langle \rho'(\boldsymbol{r},t) \rangle = 0$. In other words, $\rho'$ and $p'$ are the zero-mean density and pressure perturbations associated with the oscillation, $\delta\rho$ and $\delta p$ are the fluctuations due to convection, and $\rho_0$ and $p_0$ are the density and pressure in the fluid at hydrostatic equilibrium (i.e. the values the density and pressure would have in the absence of convection and oscillation).

We will also assume that the time derivative of the oscillating quantities $\boldsymbol{u}'$, $\rho'$, and $p'$ average to zero over a time $t_{\rm osc}$, as is the case for periodic oscillations.

The Reynolds decomposition above allows the oscillation to couple to convection through the non-linear term in Navier–Stokes equation, but does not allow mode–mode coupling for $p$ modes or inertial waves. This could only be captured by having a sum of different $\boldsymbol{u}'$, each for a different wave, in the decomposition for $\boldsymbol{u}$. Since $p$ modes are excited by the turbulent convective flow itself, modes with different frequencies co-exist and interact with each other. However, $p$ mode damping through mode–mode coupling is believed to be negligible in the Sun due to the small amplitude of the modes (Kumar & Goldreich 1989; Weinberg, Arras & Pramanik 2021), and therefore this is neglected here. Similarly, a tidally excited inertial wave could in principle interact with free inertial waves if these were present in the flow. However, numerical simulations of Jupiter have found that free inertial waves could not be maintained, probably because of their interaction with convection and gravity waves (Glatzmaier 2018). Although the question of whether these waves are present or not is still open, we will neglect the possibility of them coupling with tidally excited modes. Our analysis therefore does not include possible resonances. Note that we also neglect possible interactions of tidal oscillations or $p$ modes with magnetic waves. If these mechanisms are a source of damping, the corresponding terms can be added to the terms we include in the analysis presented below.

The kinetic energy of the flow per unit volume is $e_k = (\rho_0 + \delta\rho + \rho')(\boldsymbol{V} + \boldsymbol{u}')(\boldsymbol{V} + \boldsymbol{u}')/2$. Averaged over $t_{\rm osc}$, this gives

$$\langle e_k \rangle = E_k + \langle e'_k \rangle + \langle \rho' \boldsymbol{u}' \rangle \cdot \boldsymbol{V}, \tag{9}$$

where we have defined $E_k = (\rho_0 + \delta\rho) V^2/2$ and $e'_k = (\rho_0 + \delta\rho + \rho') u'^2/2$.

### 3.2 Equations for the kinetic energy averaged over the oscillation time-scale

As will be discussed in Section 3.9, neglecting density perturbations in Navier–Stokes equation before performing time averages to derive equations for the oscillation and the mean flow yields inconsistencies. Therefore, we first derive the equations for the conservation of kinetic energy without making any approximations. The small density perturbation limit will then be examined in the next subsection.

Substituting the Reynolds decomposition in Navier–Stokes equation (2) and averaging over $t_{\rm osc}$ yields

$$(\rho_0 + \delta\rho)\frac{\partial V_i}{\partial t} + \left\langle \rho'\frac{\partial u'_i}{\partial t}\right\rangle + (\rho_0 + \delta\rho)V_j\frac{\partial V_i}{\partial x_j} + (\rho_0 + \delta\rho)\left\langle u'_j\frac{\partial u'_i}{\partial x_j}\right\rangle + \left\langle \rho'\frac{\partial u'_i}{\partial x_j}\right\rangle V_j + \left\langle \rho' u'_j\right\rangle\frac{\partial V_i}{\partial x_j} + \left\langle \rho' u'_j\frac{\partial u'_i}{\partial x_j}\right\rangle =$$
$$-\frac{\partial}{\partial x_i}(p_0 + \delta p) + (\rho_0 + \delta\rho)g_i - 2(\rho_0 + \delta\rho)\boldsymbol{\Omega}\times\boldsymbol{V}|_i - 2\left\langle \rho'\boldsymbol{\Omega}\times\boldsymbol{u}'\right\rangle|_i + \left\langle \rho' f_{t,i}\right\rangle + \frac{\partial \left\langle \sigma_{ij}\right\rangle}{\partial x_j}, \quad (10)$$

where $|_i$ denotes the $i$ component. To derive the above equation, we have assumed that $\boldsymbol{g}$ and $\boldsymbol{\Omega}$ are not affected by the perturbations. We have also interchanged time averages with space derivatives. Note that $p_0$ and $\rho_0$ satisfy the hydrostatic equilibrium equation $-\partial p_0/\partial x_i + \rho_0 g_i = 0$. This could be subtracted off, but we keep these terms as it makes the discussion about energies more clear.

Averaging the mass conservation equation (1) over $t_{\rm osc}$ yields

$$\frac{\partial(\delta\rho)}{\partial t} + \frac{\partial}{\partial x_j}\left[(\rho_0 + \delta\rho)V_j + \left\langle \rho' u'_j\right\rangle\right] = 0, \quad (11)$$

and subtracting from equation (1) gives

$$\frac{\partial \rho'}{\partial t} + \frac{\partial}{\partial x_j}\left[(\rho_0 + \delta\rho)u'_j + \rho' V_j + \rho' u'_j - \left\langle \rho' u'_j\right\rangle\right] = 0. \quad (12)$$

We obtain an equation for $E_k$ by multiplying equation (10) by $V_i$ and summing over $i$. Using equations (11) and (12) then yields

$$\frac{\partial E_k}{\partial t} = -\frac{\partial}{\partial x_j}\left[(E_k + p_0 + \delta p)V_j + (\rho_0 + \delta\rho)V_i\left\langle u'_i u'_j\right\rangle + \frac{1}{2}V^2\left\langle \rho' u'_j\right\rangle + V_i V_j\left\langle \rho' u'_i\right\rangle + V_i\left\langle \rho' u'_i u'_j\right\rangle - \left\langle \sigma_{ij}\right\rangle V_i\right] - V_i\left\langle \frac{\partial}{\partial t}(\rho' u'_i)\right\rangle$$
$$+\mathcal{D}_R - \mathcal{D}_v + (\rho_0 + \delta\rho)\boldsymbol{g}\cdot\boldsymbol{V} + (p_0 + \delta p)\nabla\cdot\boldsymbol{V} - 2\left\langle \rho'\boldsymbol{\Omega}\times\boldsymbol{u}'\right\rangle\cdot\boldsymbol{V} + \left\langle \rho'\boldsymbol{f}_t\right\rangle\cdot\boldsymbol{V}, \quad (13)$$

where we have defined

$$\mathcal{D}_R = \left[(\rho_0 + \delta\rho)\left\langle u'_i u'_j\right\rangle + \left\langle \rho' u'_i u'_j\right\rangle + \left\langle \rho' u'_i\right\rangle V_j\right]\frac{\partial V_i}{\partial x_j}, \quad (14)$$

$$\mathcal{D}_v = \left\langle \sigma_{ij}\right\rangle\frac{\partial V_i}{\partial x_j}. \quad (15)$$

Similarly, we obtain an equation for $e'_k$ by multiplying equation (2) by $u'_i$, summing over $i$, using equation (1) and averaging over $t_{\rm osc}$. This yields

$$\left\langle \frac{\partial e'_k}{\partial t}\right\rangle = -\frac{\partial}{\partial x_j}\left[\left\langle e'_k\right\rangle V_j + \left\langle e'_k u'_j\right\rangle + \left\langle p' u'_j\right\rangle - \left\langle \sigma_{ij} u'_i\right\rangle\right] - \left\langle \rho' u'_i\right\rangle\frac{\partial V_i}{\partial t}$$
$$-\mathcal{D}_R - \mathcal{D}'_v + \left\langle \rho'\boldsymbol{g}\cdot\boldsymbol{u}'\right\rangle + \left\langle p'\nabla\cdot\boldsymbol{u}'\right\rangle + 2\left\langle \rho'\boldsymbol{\Omega}\times\boldsymbol{u}'\right\rangle\cdot\boldsymbol{V} + \left\langle (\rho_0 + \delta\rho + \rho')\boldsymbol{f}_t\cdot\boldsymbol{u}'\right\rangle, \quad (16)$$

where

$$\mathcal{D}'_v = \left\langle \sigma_{ij}\frac{\partial u'_i}{\partial x_j}\right\rangle. \quad (17)$$

No approximations have been made to obtain the equations above, i.e all the terms have been retained. As pointed out above, these equations apply whether the oscillation is an equilibrium tide, the sum of an equilibrium tide and a propagating inertial wave, or a propagating pressure wave (in which case $\boldsymbol{f}_t$ has to be thought of as the force that excites the wave). They also apply even if the oscillations are not periodic, in which case the average has to be taken over a time long compared to $t_{\rm osc}$ and short compared to $t_{\rm conv}$. Therefore, these equations can be used when the role of the tidal oscillations and that of convection are reversed, i.e. when the tides are the mean flow and convective motions are the rapid fluctuations, which is appropriate when $t_{\rm conv} \ll t_{\rm osc}$ (note, however, that in that case the fluctuating velocity does not average to zero, but to a value which is second order in the fluctuations, e.g. Nordlund, Stein & Asplund 2009).

### 3.3 The case of small perturbations

We now approximate the energy conservation equations (13) and (16) using $|\rho'| \ll \rho_0$ and $|\boldsymbol{u}'| \ll |\boldsymbol{V}|$, which is always satisfied for tidal oscillations or pressure modes in a convective flow.

The mass conservation equation (12) can then be approximated as

$$\frac{\partial \rho'}{\partial t} + \frac{\partial}{\partial x_j}\left[(\rho_0 + \delta\rho)u'_j + \rho' V_j\right] = 0. \quad (18)$$

We note $\lambda_{\rm conv}$ the characteristic spatial scale of the convective eddies (mixing length), such that $V \sim \lambda_{\rm conv}/t_{\rm conv}$ (if $V$ includes significant contribution from differential rotation, then $t_{\rm conv}$ and $\lambda_{\rm conv}$ are themselves affected by rotation). We further note $\lambda_{\rm osc}$ the characteristic spatial scale over which the oscillating quantities vary. For the equilibrium tide in Jupiter, Saturn, or the Sun, and in the parts of the envelope where tides are significant, $\lambda_{\rm osc}/\lambda_{\rm conv} \gg 1$ near the surface and decreases to reach values on the order of unity deeper in the envelope. Since inertial waves are driven by the Coriolis force acting on the equilibrium tide (Ogilvie 2013), their $\lambda_{\rm osc}$ is comparable to that of the equilibrium tide. For $p$ modes, $\lambda_{\rm osc}$ is larger than the pressure scale height $H_p$. Since $\lambda_{\rm conv} \sim 2H_p$, it follows that $\lambda_{\rm conv} \lesssim \lambda_{\rm osc}$ in all cases.

Therefore, $\left|\partial \left(\rho' V_j\right)/\partial x_j\right| \sim |\rho'| V/\lambda_{\rm conv} \sim |\rho'|/t_{\rm conv}$. Since $t_{\rm osc} \ll t_{\rm conv}$, this is very small compared to $|\partial \rho'/\partial t| \sim |\rho'|/t_{\rm osc}$, so that equation (18) can be further approximated as:

$$\frac{\partial \rho'}{\partial t} + \frac{\partial}{\partial x_j}\left[(\rho_0 + \delta\rho) u'_j\right] = 0. \tag{19}$$

This yields $|\rho'|/t_{\rm osc} \sim \left|\partial \left(\rho_0 u'_j\right)/\partial x_j\right| \sim \rho_0 u'/\lambda_{\rm osc}$ or $\sim \rho_0 u'/r$, depending on whether the oscillation is compressible or not. This implies

$$\epsilon \equiv \frac{|\rho'|}{\rho_0}\frac{V}{u'} \lesssim \frac{\lambda_{\rm conv}}{\lambda_{\rm osc}}\frac{t_{\rm osc}}{t_{\rm conv}} \ll 1. \tag{20}$$

Therefore, in equation (9), $\left|\langle \rho' u' \rangle \cdot V\right|/\left|\langle e'_k \rangle\right| \sim \epsilon \ll 1$. The kinetic energy averaged over the tidal period is then $\langle e_k \rangle \simeq E_k + \langle e'_k \rangle$, which is the sum of the kinetic energy of the mean flow and that of the oscillation.

Furthermore, having $\epsilon \ll 1$ and $|\rho'| \ll \rho_0$ yields $\mathcal{D}_R \simeq \rho_0 D_R$ (where we have also used $|\delta\rho| \ll \rho_0$), with

$$D_R = \langle u'_i u'_j \rangle \frac{\partial V_i}{\partial x_j}, \tag{21}$$

which is the parameter that was first introduced in Terquem (2021).

### 3.4 Averaging over the convective time-scale

As we are interested in the exchange of energy between convection and the oscillation over a time-scale long compared to the convective time-scale, we now average equations (13) and (16) over such a time-scale.

The exact same procedure is followed when the mixing length approximation is used, which may be appropriate when $t_{\rm conv} \ll t_{\rm osc}$, and which corresponds to the tidal oscillation being the mean flow and convection being the rapid fluctuations. In that case, the first time averaging is done over the convective time-scale, and the second over the oscillation period. Mixing length theory assumes that $D_R$, which involves a coupling between the convective Reynolds stress and the gradient of the tidal velocity, is always negative, corresponding to local dissipation of the tides (mean flow). Therefore, $D_R$ does not average to zero over an oscillation period, even though it is linear in the gradient of the tidal velocity (see Appendix A for a more detailed discussion).

Similarly, here, where $t_{\rm conv} \gg t_{\rm osc}$ and $D_R$ involves a coupling between the Reynolds stress of the oscillation and the gradient of the convective velocity, we allow for the possibility that the oscillation is locally dissipated. This corresponds to $D_R > 0$ and implies that $\overline{D_R} \ne 0$, where the overline denotes an average over a time long compared to the convective time-scale.

Locally, $|D_R|$ given by equation (21) is of order $u'^2 V/\lambda_{\rm conv} \sim u'^2/t_{\rm conv}$. If the oscillation is *locally* dissipated, then $D_R$ is positive everywhere and at all times, and $\overline{D_R} = |D_R|$. However, if the oscillation is not locally dissipated, $\left|\overline{D_R}\right|$ may be much smaller than $|D_R|$, in which case the rate of energy dissipation is too low to explain the circularization of solar-type binaries (Terquem & Martin 2021). We will neglect in equations (13) and (16) the terms which are small compared to $\rho_0 |D_R|$ when averaged over a long time-scale, while retaining $\rho_0 D_R$ to allow for the possibility of local dissipation. If $\left|\overline{D_R}\right|$ is actually small compared to $|D_R|$, then the exchange of energy between the oscillation and convective eddies with long time-scale is negligible and cannot account for the observations, in which case the terms that we neglect are not important anyway.

We therefore neglect $\langle \rho' u'_i \rangle (\partial V_i/\partial t)$ in equation (16), as it is $\epsilon$ times smaller than $\rho_0 |D_R|$. We also note that, in equation (13), $\langle \partial \left(\rho' u'_i\right)/\partial t \rangle = 0$ because $\rho' u'_i$ is the sum of constant and periodic terms.

We now compare the Coriolis term, which redistributes kinetic energy among the different components of the velocities, to $\rho_0 |D_R|$. We have $\left|\langle \rho' \boldsymbol{\Omega} \times u' \rangle \cdot V\right| \sim |\rho'| u' \Omega V$, as $\rho'$ is almost in phase with $u'_\varphi$ (they would be exactly in phase if there were no exchange of energy between the oscillation and the convective flow). The ratio of this quantity to $\rho_0|D_R|$ is $(\lambda_{\rm conv}/\lambda_{\rm osc}) \Omega t_{\rm osc}$. If $\Omega$ is large, then $\Omega t_{\rm osc} \simeq \pi$, so that this ratio is of order unity. However, mass conservation implies that $\overline{V} \sim (|\delta\rho|/\rho_0) V \ll V$ (Nordlund et al. 2009). Therefore the Coriolis term averaged over a long time-scale is very small compared to $\rho_0|D_R|$, and will therefore be neglected.

The work done by the tidal force on the convective flow is given by the last term on the right-hand side of equation (13). Averaged over a long time-scale, this is $\langle \rho' \boldsymbol{f}_t \rangle \cdot \boldsymbol{V} \sim |\rho'| f_t \overline{V}$. The work done by the tidal force on the oscillation is given by the last term on the right-hand side of equation (16), and is $\langle \rho \boldsymbol{f}_t \cdot \boldsymbol{u}' \rangle \sim \rho_0 f_t u' \delta$, where $\delta$ is the (small) phase shift between the tidal displacement and the tidal force which results from the exchange of energy between the oscillation and convection. The ratio of these two terms is therefore $\eta \equiv |\rho'|\overline{V}/(\rho_0 u' \delta) = \epsilon \overline{V}/(V\delta)$. The work done by the buoyancy force $g|\delta\rho|$ over $\sim \lambda_{\rm conv}$ is equal to the kinetic energy per unit volume of convective motions (Schwarzschild 1958), so that $|\delta\rho|/\rho_0 \sim V^2/(g\lambda_{\rm conv}) \sim V/(gt_{\rm conv})$, and this is equal to $\overline{V}/V$. Therefore, $\eta \sim \epsilon V/(gt_{\rm conv}\delta)$. We will show in Section 5.1 that $\eta \ll 1$ for tidal oscillations, so that the last term on the right-hand side of equation (13) can be neglected: most of the tidal work is done on the oscillating velocities, not on the mean flow. This implies that the mean flow 'feels' the effect of the tidal forcing through exchanging energy with the oscillation, rather than directly. (For *p* modes, there is no work done by the forcing on the convective flow, as the forcing comes from convection itself).

Finally, we note that $\overline{V} \ll V$ also implies that the term $\partial \left(\langle e'_k \rangle V_j\right)/\partial x_j$ averaged over a long time-scale is very small compared to $\rho_0|D_R|$ (we are assuming that the gradient of the convective velocity, not the velocity itself, may couple to the tidal Reynolds stress).

Using the small perturbation approximations described in Section 3.3 and averaging equations (13) and (16) over a time long compared to the convective time-scale while retaining $D_R$ then yields

$$\frac{\partial E_k}{\partial t} = -\frac{\partial}{\partial x_j}\left[(E_k + p_0 + \delta p)V_j - \langle\sigma_{ij}\rangle V_i\right] + \rho_0 D_R - \mathcal{D}_v + (\rho_0 + \delta\rho)\,\mathbf{g}\cdot\mathbf{V} + (p_0 + \delta p)\,\nabla\cdot\mathbf{V}, \quad (22)$$

$$\left\langle\frac{\partial e'_k}{\partial t}\right\rangle = -\frac{\partial}{\partial x_j}\left(\langle p'u'_j\rangle - \langle\sigma_{ij}u'_i\rangle\right) - \rho_0 D_R - \mathcal{D}'_v + \langle\rho'\mathbf{g}\cdot\mathbf{u}'\rangle + \langle p'\nabla\cdot\mathbf{u}'\rangle + \rho_0\langle\mathbf{f}_t\cdot\mathbf{u}'\rangle, \quad (23)$$

where the average over the long time-scale is now taken as read. Equations (22) and (23) show that kinetic energy is exchanged between the mean flow and the oscillation at a rate $\rho_0 D_R$ per unit volume, which extends the result that was established by Terquem (2021) in the incompressible and non–rotating case.

### 3.5 Equations for the potential and internal energy

We now write the equations for the potential and internal energy of the oscillation and mean flow to identify the terms which exchange the kinetic energy of the oscillation with other forms of energy.

The potential energy per unit volume is $e_p = (\rho_0 + \delta\rho + \rho')(\Psi_0 + \Psi_t)$. Averaged over $t_{\rm osc}$, we get $\langle e_p\rangle = E_p + \langle e'_p\rangle$ with $E_p = (\rho_0 + \delta\rho)\Psi_0$ being the potential energy of the mean flow and $\langle e'_p\rangle = \langle\rho'\Psi_t\rangle$ being the mean potential energy of the oscillation.

We obtain an equation for $E_p$ by multiplying equation (11) by $\Psi_0$, which yields:

$$\frac{\partial E_p}{\partial t} = -\frac{\partial}{\partial x_j}\left(E_p V_j + \langle\rho'\Psi_0 u'_j\rangle\right) - (\rho_0 + \delta\rho)\,\mathbf{g}\cdot\mathbf{V} - \langle\rho'\mathbf{g}\cdot\mathbf{u}'\rangle, \quad (24)$$

where we have used the fact that $\Psi_0$ and $\rho_0$ are independent of time.

A conservation equation for $e'_p$ is obtained by multiplying equation (1) by $\Psi_t$ and averaging over $t_{\rm osc}$:

$$\left\langle\frac{\partial e'_p}{\partial t}\right\rangle = -\frac{\partial}{\partial x_j}\left[\langle e'_p\rangle V_j + (\rho_0 + \delta\rho)\langle\Psi_t u'_j\rangle\right] + \left\langle\rho'\frac{\partial\Psi_t}{\partial t}\right\rangle - \rho_0\langle\mathbf{f}_t\cdot\mathbf{u}'\rangle, \quad (25)$$

where we have used $\eta \ll 1$.

For a perfect gas, the internal energy per unit volume is $e_{\rm int} = p/(\gamma - 1)$, where $\gamma$ is the ratio of the heat capacity at constant pressure to that at constant volume. Assuming this parameter not to be affected by the perturbation, we have $\langle p'/(\gamma - 1)\rangle = 0$ and $\langle e_{\rm int}\rangle = (p_0 + \delta p)/(\gamma - 1)$ is then the internal energy of the mean flow: on average over a tidal period, there is no internal energy in the tidal oscillation. A conservation equation for $e_{\rm int}$ is obtained by averaging equation (5) over $t_{\rm osc}$

$$\left\langle\frac{\partial e_{\rm int}}{\partial t}\right\rangle = \frac{\partial}{\partial t}\left(\frac{p_0 + \delta p}{\gamma - 1}\right) = -\frac{\partial}{\partial x_j}\left[\frac{(p_0 + \delta p)V_j}{\gamma - 1} + \frac{\langle p'u'_j\rangle}{\gamma - 1}\right] - (p_0 + \delta p)\,\nabla\cdot\mathbf{V} - \langle p'\nabla\cdot\mathbf{u}'\rangle + \mathcal{D}_v + \mathcal{D}'_v - \langle\nabla\cdot\mathbf{q}\rangle. \quad (26)$$

### 3.6 Identifying all the terms responsible for energy transfer between oscillation and mean flow

As can be seen from equation (22) to (26), to leading order, the work done by the tidal force enters through the kinetic and potential energies of the oscillation only. The potential energy of the oscillation cannot be exchanged with the mean flow. However, the kinetic energy of the oscillation can be exchanged with the kinetic energy of the convective flow through the term $\rho_0 D_R$, with the potential energy of the convective flow through the buoyancy term $\langle\rho'\mathbf{g}\cdot\mathbf{u}'\rangle$, and with the internal energy of the convective flow through the pressure term $\langle p'\nabla\cdot\mathbf{u}'\rangle$ and viscous term $\mathcal{D}'_v$.

The compressibility associated with low-frequency tidal oscillations is negligible. This can be seen by comparing $t_{\rm osc}$ with the time $t_s = \lambda_{\rm osc}/c_s$ it takes a sound wave to cross the characteristic spatial scale of the oscillation, where $c_s$ is the sound speed. We have $c_s \sim \sqrt{p_0/\rho_0} \sim \sqrt{gH_p}$, where $H_p$ is the pressure-scale height, so that $t_s \sim (\lambda_{\rm osc}/\sqrt{rH_p})\omega_0^{-1}$, where $\omega_0 \equiv \sqrt{g/r}$ is the natural frequency. The perturbation is therefore approximately incompressible if $t_s/t_{\rm osc} \sim (\lambda_{\rm osc}/\sqrt{rH_p})(\omega_0 t_{\rm osc})^{-1} \ll 1$. We have checked that this is satisfied for the tidal periods of interest in Jupiter, Saturn, and the Sun, although it is only marginally satisfied very close to the surface in Jupiter and Saturn. Therefore, the effect of compressibility on tidal oscillations can be neglected (Lighthill 1978), which means that terms involving $\nabla\cdot\mathbf{u}'$ (and therefore also $\nabla\cdot\boldsymbol{\xi}$, where $\boldsymbol{\xi}$ is the Lagrangian tidal displacement) can be discarded.

The buoyancy term is $\langle\rho'\mathbf{g}\cdot\mathbf{u}'\rangle \simeq \langle\rho' g u'_r\rangle$, since gravity is mostly in the radial direction. The mass conservation equation (19) can be written as $\rho' = -\nabla\cdot(\rho_0\boldsymbol{\xi})$, where we have used $|\delta\rho| \ll \rho_0$. This shows that $\langle\rho' u'_r\rangle = -\rho_0\langle u'_r\nabla\cdot\boldsymbol{\xi}\rangle$, as $\rho_0$ varies with $r$ only.

Therefore, both the pressure and buoyancy terms can be neglected for tidal oscillations. Note that they were also neglected compared to the term involving the Reynolds stress in the study of gravity waves or $f$-modes interacting with convection in Press (1981), Goldreich & Kumar (1990), Lecoanet & Quataert (2013).

For $p$ modes however, $t_s/t_{\rm osc}$ becomes of order unity near the surface of the convective zone of the Sun, and compressibility therefore plays a role for the damping of these modes. This yields a phase shift between $\rho'$ and $p'$ which results in net work done by the pressure force (Samadi, Belkacem & Sonoi 2015, see also Goldreich & Kumar 1990 for a comparison of these terms with the term involving the Reynolds stress).

The viscous flux and $\mathcal{D}'_v$ in equation (23) can also be neglected, because microscopic viscosity has a negligible effect on oscillations, except possibly near the surfaces of the envelope. In general, viscous dissipation is negligible away from the surfaces of the convective envelope, both in the Sun (Miesch 2005) and in Jupiter (Guillot et al. 2004). In the bulk of the convective zone, the energy contained in the large scale eddies is transported towards the surface mostly by the enthalpy flux, with comparatively very little dissipation due to viscous stresses acting on the smaller scales. Near the surface, the enthalpy flux falls off and energy is transported by the radiative flux (e.g. Featherstone & Miesch 2015).

### 3.7 Tidal work and dissipation

We now consider the specific case of a tidal oscillation, for which the terms related to buoyancy, compressibility and viscosity can be neglected in equation (23). The divergence term in this equation can also be neglected, because $\langle p' \boldsymbol{u}' \rangle$ is associated with compressibility, which we have just shown is not playing any role.

Equation (23) can therefore be written as

$$\left\langle \frac{\partial e_k'}{\partial t} \right\rangle = -\rho_0 D_R + \rho_0 \langle \boldsymbol{f}_t \cdot \boldsymbol{u}' \rangle. \tag{27}$$

Equations (25) and (27) have been obtained after averaging over a time long compared to the convective time-scale, so that the time derivatives on the left-hand side are changes over long time-scales. Since the brackets are time averages over short time-scales, they can be swapped with the time derivatives, i.e. $\langle \partial e_k' / \partial t \rangle = \partial \langle e_k' \rangle / \partial t$ and similarly for $e_p'$. Therefore, given that $\langle e_k' \rangle$ and $\langle e_p' \rangle$ are constant, equations (25) and (27) become

$$\rho_0 \langle \boldsymbol{f}_t \cdot \boldsymbol{u}' \rangle = -\frac{\partial}{\partial x_j} \left( \langle e_p' \rangle V_j + \rho_0 \langle \Psi_t u_j' \rangle \right) + \left\langle \rho' \frac{\partial \Psi_t}{\partial t} \right\rangle, \tag{28}$$

$$\rho_0 \langle \boldsymbol{f}_t \cdot \boldsymbol{u}' \rangle = \rho_0 D_R. \tag{29}$$

Integrating equation (28) over the volume $\mathcal{V}$ of the convective envelope yields

$$\int_{\mathcal{V}} \rho_0 \langle \boldsymbol{f}_t \cdot \boldsymbol{u}' \rangle \, \mathrm{d}v = \int_{\mathcal{V}} \left\langle \rho' \frac{\partial \Psi_t}{\partial t} \right\rangle \, \mathrm{d}v, \tag{30}$$

where we have used the fact that $\rho_0$ is very small on the outer surface and $\boldsymbol{u}'$ is very small on the inner surface (as in Section 2.2), so that the divergence term does not contribute. Comparing this equation with equation (8) shows that the energy which is ultimately radiated away comes from the work done by the tidal force (when this is positive).

We now discuss the implications of equation (29).

When there is no exchange of energy between the tidal flow and convection, i.e. $D_R = 0$, the tidal displacement $\boldsymbol{\xi}$ is in phase with the tidal force $\boldsymbol{f}_t$, so that $\boldsymbol{u}' = \partial \boldsymbol{\xi}/\partial t$ is in quadrature and $\langle \boldsymbol{f}_t \cdot \boldsymbol{u}' \rangle = 0$. This means that $\boldsymbol{f}_t \cdot \boldsymbol{u}'$ is positive during half a tidal period and negative during the other half. When it is positive, the kinetic energy of the oscillation increases while its potential energy decreases by the same amount. When it is negative, the kinetic energy decreases and the potential energy increases.

When there is dissipation, i.e. $D_R > 0$, $\boldsymbol{\xi}$ lags behind $\boldsymbol{f}_t$ by a (positive) phase shift $\delta$, which yields $\langle \boldsymbol{f}_t \cdot \boldsymbol{u}' \rangle \propto \sin \delta$ with $\langle \boldsymbol{f}_t \cdot \boldsymbol{u}' \rangle > 0$. In other words, when the tidal flow transfers energy to convection, work is done by the tidal force, which supplies the energy being transferred. This ultimately comes from the orbital motion of the binary.

If $D_R < 0$ instead, the tidal flow extracts kinetic energy from convection. In that case, $\boldsymbol{\xi}$ leads $\boldsymbol{f}_t$ and the situation is as above but with $\delta < 0$. The work done by the tidal force on the flow is now negative. In other words, the energy extracted from convection by the tidal oscillation is removed from the flow by the work done by the tidal force, and is ultimately fed to the orbital motion of the binary.

The analysis above shows that tidal energy cannot be dissipated by the oscillation itself, as it cannot be converted into thermal energy of the oscillation. It can only be dissipated by being transferred to the kinetic energy of the mean convective flow first, where it becomes part of the convective energy budget and is transformed into thermal energy in the standard way.

### 3.8 Equilibrium tide and inertial waves

As already mentioned above, when the tidal frequency in the rotating frame is less than twice the rotational frequency, propagating inertial waves are excited. This corresponds to $t_{\mathrm{orb}} \geq t_{\mathrm{rot}}/2$, where $t_{\mathrm{rot}}$ is the rotational period.

When this condition is satisfied, the oscillation is the sum of a non-wave like part and an inertial wave. The non-wave like part is the equilibrium tide, which corresponds to the flow being always instantaneously at hydrostatic equilibrium in the perturbed potential, and is therefore a solution of the equations without the Coriolis force. The Coriolis force acting on the equilibrium tide is then a forcing term that drives inertial waves (see Ogilvie 2013 for a thorough discussion). Denoting $\boldsymbol{u}_e'$ and $\boldsymbol{u}_w'$ the velocities associated with the equilibrium tide and inertial waves, respectively, we then have $|\boldsymbol{u}_e'|/|\boldsymbol{u}_w'| \sim |\omega_{\mathrm{osc}}|/(2\Omega) \leq 1$. When inertial waves are present, $D_R$ can be written as $D_{R,e} + D_{R,w} + D_{R,ew}$, with

$$D_{R,e} = \langle u_{e,i}' u_{e,j}' \rangle \frac{\partial V_i}{\partial x_j}, \quad D_{R,w} = \langle u_{w,i}' u_{w,j}' \rangle \frac{\partial V_i}{\partial x_j}, \quad D_{R,ew} = \left( \langle u_{e,i}' u_{w,j}' \rangle + \langle u_{e,j}' u_{w,i}' \rangle \right) \frac{\partial V_i}{\partial x_j}. \tag{31}$$

This shows that the exchange of energy between the convective flow and the tidal oscillation can be written in the same way whether the perturbation is dominated by the equilibrium tide, by inertial waves, or by both. Therefore, those inertial waves which have a period small compared to the convective time-scale can only be dissipated by interaction with convection if $D_{R,w} > 0$.

### 3.9 Comments on Barker & Astoul (2021)

Barker & Astoul (2021) have claimed that the term exchanging kinetic energy between convection and tidal oscillations is $\rho_0 D_R - t_1$ with $t_1 \equiv V_i \langle u'_i (\partial \rho'/\partial t) \rangle$ (using their notations), instead of $\rho_0 D_R$, as found above. Their result is based on identifying the term exchanging kinetic energy as being $\rho_0 \mathbf{V} \cdot (\mathbf{u}' \cdot \nabla) \mathbf{u}'$ only, i.e. neglecting the contribution from the local time derivative $\rho (\partial \mathbf{u}/\partial t)$ and also other terms in $\rho (\mathbf{u} \cdot \nabla) \mathbf{u}$ in Navier–Stokes equation. However, as they bring $\partial \rho'/\partial t$ back from the mass conservation equation, their analysis is not self-consistent and does not yield the correct energy conservation equation. Including the local time derivative would indeed add the term $\mathbf{V} \cdot \rho' (\partial \mathbf{u}'/\partial t)$ which, added to $t_1$, gives $V_i \langle \partial (\rho' u'_i) /\partial t \rangle$, which is zero (see Section 3.4).

When all the terms are taken into account in a self-consistent manner, the correct exchange term is therefore $\rho_0 D_R$, as shown in the analysis presented above.

We also note that, contrary to what is argued by Barker & Astoul (2021), the term $\rho_0 V_i V_j (\partial u'_i/\partial x_j)$, which has been regarded as the term exchanging energy between tides and convection in previous studies (as will be discussed in the next section), is not an alternative to $\rho_0 D_R$ when $t_{\rm osc} \ll t_{\rm conv}$. Indeed, $\rho_0 V_i V_j (\partial u'_i/\partial x_j)$ averages to zero over the shortest time-scale $t_{\rm osc}$. This is discussed in more details in Appendix A.

## 4 VISCOUS AND TURBULENT SHEAR FLOWS, SLOW AND FAST TIDES

We now discuss how to express $D_R$ in terms of the flow velocities. Before considering tidal oscillations in convective flows, we first review the classical treatment of viscous and turbulent shear flows, as this is instructive and this has been used to approximate tidal dissipation in the standard approach, when the tides are the mean flow. We then discuss both the standard approach and the case of fast tides to which the analysis of the previous section applies. Although, in this section, we refer to the oscillation as a tide, the discussion also applies to $p$ modes. For those modes however, $D_R$ is only part of the exchange energy rate with the convective flow.

### 4.1 Viscous and turbulent shear flows

In a viscous flow with mean velocity $\mathbf{U}$, momentum is transported in the direction of decreasing momentum by the fluctuating (thermal) part $\mathbf{c}$ of the particle velocities. After travelling through a mean free path $\lambda$, particles collide with each other and redistribute momentum. For an incompressible Newtonian fluid, the viscous stress is given by

$$\sigma_{ij} = -\rho \langle c_i c_j \rangle = \rho \nu \left( \frac{\partial U_i}{\partial x_j} + \frac{\partial U_j}{\partial x_i} \right), \qquad (32)$$

where $\nu \sim c\lambda/3$ is the kinematic viscosity. This corresponds to a rate of change of energy per unit mass for the mean flow given by

$$D_{R,\rm visc} = \langle c_i c_j \rangle \frac{\partial U_i}{\partial x_j} = -\frac{1}{2} \nu \left( \frac{\partial U_i}{\partial x_j} + \frac{\partial U_j}{\partial x_i} \right)^2. \qquad (33)$$

As $D_{R,\rm visc} < 0$, kinetic energy is irreversibly *lost* by the mean flow. In this case, the correlations $<c_i c_j>$ between the components of the velocity fluctuations are a *result* of the shear, and are given the sign required for kinetic energy to be transformed into thermal energy.

In the classical case of a turbulent shear flow, as described, e.g., in Tennekes & Lumley (1972), the Reynolds stress $-\rho <c_i c_j>$ is associated with the velocity $\mathbf{c}$ of the turbulent fluctuations. The rate of change of energy per unit mass for the mean shear flow, which has velocity $\mathbf{U}$, is still $D_{R,\rm turb} = <c_i c_j> (\partial U_i/\partial x_j)$. Because the length scale of the turbulent eddies is small compared to the scale of the shear flow, eddies are stretched by the shear flow, and conservation of angular momentum then produces a correlation of the components of the turbulent velocity yielding $D_{R,\rm turb} < 0$. Therefore, here again, $<c_i c_j>$ is determined by the shear. This corresponds to a transfer of energy from the mean flow to the largest turbulent eddies and the subsequent cascade results in a small scale viscous dissipation of the free energy present in the shear flow. This is consistent with the fact that the turbulence is due to instabilities of the mean shear flow itself, so that the energy of the turbulent eddies comes from the shear.

We now discuss energy transfer in convective flows subject to tidal oscillations.

### 4.2 Slow tides $t_{\rm conv} \ll t_{\rm osc}$ and mixing length theory

In the analysis done in the previous section, it was assumed that the tidal period was the shortest time-scale. However, the same analysis could be done for the case where the convective time-scale is the shortest time-scale. *Convective motions would then be the rapid fluctuations, whereas the tidal oscillation would be the mean flow.* This is the standard picture which has been considered by previous authors. This yields a rate of change of kinetic energy which is still given by

$$D_{R,\rm slow} = \langle u'_i u'_j \rangle \frac{\partial V_i}{\partial x_j}, \qquad (34)$$

but *with V being the tidal velocity and u′ being the convective velocity*, and the average is over a time long compared to the convective time-scale and short compared to the tidal period (see Appendix A). Since the mean flow is now the tidal oscillation, transfer of energy from the tides to convection requires $D_{R,\,\text{slow}} < 0$.

In previous studies, starting with Zahn (1966), the mixing length approximation has been used to write $D_{R,\,\text{slow}}$ in terms of a turbulent convective viscosity. This relies on assuming that convective eddies behave like particles in a fluid, and exchange momentum with their environment when they dissolve after having travelled over a mixing length $\lambda_{\text{conv}}$, which is (at most) twice the pressure scale height. The Reynolds stress, $-\rho \langle u_i' u_j' \rangle$, is then expressed by analogy with the viscous stress (32) as

$$-\rho \langle u_i' u_j' \rangle = \rho \nu_t \left( \frac{\partial V_i}{\partial x_j} + \frac{\partial V_j}{\partial x_i} \right), \quad (35)$$

where $\nu_t \sim u' \lambda_{\text{conv}}$ is the turbulent convective viscosity. The rate of energy dissipation is then given by equation (33) with $\nu$ being replaced by $\nu_t$, which yields

$$\left| D_{R,\,\text{slow}} \right| \sim u' \lambda_{\text{conv}} \left( \frac{V}{\lambda_{\text{osc}}} \right)^2 \sim \frac{V^2}{t_{\text{conv}}} \left( \frac{\lambda_{\text{conv}}}{\lambda_{\text{osc}}} \right)^2, \quad (36)$$

where we have used $|\partial V_i / \partial x_j| \sim V/\lambda_{\text{osc}}$ and $u' \sim \lambda_{\text{conv}}/t_{\text{conv}}$. Therefore, $D_{R,\,\text{slow}}$ is linear in the convective velocity $u'$ and quadratic in the tidal velocity $V$.

The mixing length approximation *assumes* that convective eddies always extract energy from the background shear flow, i.e. transport the momentum associated with the tides from regions where it is high to regions where it is lower. In other words, in this picture, the tidal oscillation dictates the direction in which convective eddies travel, yielding the correlations $\langle u_i' u_j' \rangle$ between the components of the convective velocity to have the sign required for tides to be dissipated. Note that, as the tidal velocity changes sign periodically, $\langle u_i' u_j' \rangle$ has to change sign on the same time-scale to keep $D_{R,\,\text{slow}} < 0$.

There are two issues with this picture. First, it is difficult to envision how tidal velocities, which are orders of magnitude smaller than convective velocities in the flows of interest, could influence the convective velocities. Second, even if the tides could affect convective motions, it is not clear this would result in the correlations $\langle u_i' u_j' \rangle$ having the required amplitude and the required sign. Indeed, although convection does correlate velocity and density fluctuations very effectively (yielding positive work from the buoyancy force), it does not necessarily produce correlations of the components of the flow velocity (Tennekes & Lumley 1972, Section 3.4). And when such correlations are produced, they do not necessarily have the sign required for convection to act as a viscosity (Starr 1968). The fact that differential rotation in the Sun is *produced* by the convective Reynolds stress is a clear example of convection acting as a negative viscosity.

Finally, we comment on the fact that the mixing length theory is based on a diffusion approximation which is not formally valid when the length scale over which the fluctuations vary is comparable to that over which the mean flow varies, which is the case when considering tides in a convective flow. This is even more of a problem in the presence of rotation, which affects the motion of the convective eddies as they move over a mixing length. In such a case, the motion of convective eddies cannot be treated in a similar way as the motion of molecules, for which rotation is irrelevant as they move over a mean free path.

There are therefore unjustified assumptions behind the model of convection acting as a turbulent viscosity. However, it gives theoretical dissipation rates which are in agreement with observed circularization periods for binaries for which $t_{\text{conv}} \ll t_{\text{osc}}$ (Verbunt & Phinney 1995).

### 4.3 Fast tides $t_{\text{conv}} \gg t_{\text{osc}}$

Here, $D_R$ is given by equation (21), where $u'$ is the tidal velocity (fluctuations) and $V$ is the convective velocity (mean flow). To leading order, $u_r'$ and $u_\theta'$ are in phase (any phase shift would be a result of energy exchange between the oscillation and the convective flow), so that $\langle u_r' u_\theta' \rangle \sim u'^2$. We also have $\langle u_r'^2 \rangle \sim \langle u_\theta'^2 \rangle \sim \langle u_\varphi'^2 \rangle \sim u'^2$, so that

$$|D_R| \sim u'^2 \frac{V}{\lambda_{\text{conv}}} \sim \frac{u'^2}{t_{\text{conv}}}, \quad (37)$$

where we have used $|\partial V_i/\partial x_j| \sim V/\lambda_{\text{conv}} \sim 1/t_{\text{conv}}$. Therefore, here again, $D_R$ is linear in the convective velocity $V$ and quadratic in the tidal velocity $u'$. This is consistent with equation (29), as the amplitude of the tidal velocity is proportional to that of the tidal force, so that the dependence on $u'$ is quadratic. The dependence on the convective velocity comes from the phase shift between the tidal force and the tidal displacement, which is due to the interaction between the tides and convection.

In previous studies, it has been argued that mixing-length theory is applicable to both the case of slow and fast tides, but that in the regime of fast tides the turbulent viscosity has to be reduced (Zahn 1966; Goldreich & Nicholson 1977). The results above show that mixing-length theory does not apply to fast tides, but none the less the tidal dissipation rate has the same form as that given by this theory, except *without the reduction factor*. The dissipation rate is actually larger than for slow tides, as $|D_{R,\,\text{slow}}/D_R| \sim (\lambda_{\text{conv}}/\lambda_{\text{osc}})^2$ (remembering that the roles of the tidal and convective velocities have been exchanged in the expression for $D_{R,\,\text{slow}}$). This is much smaller than unity in the parts of the convective envelopes of the Sun, Jupiter, and Saturn which contribute to tidal dissipation, which yields enhanced exchange of energy when $t_{\text{conv}} \gg t_{\text{osc}}$.

In general, at a given location in a convective flow, there is a range of convective time-scales associated with eddies of different sizes. Therefore, even when $t_{\text{osc}}$ is small compared to the longest convective time-scale, it is still likely to be large compared to the shortest one. This implies that values of $t_{\text{osc}}/t_{\text{conv}}$ smaller and larger than unity simultaneously contribute to exchange of energy. However, since $|D_{R,\,\text{slow}}|$ is significantly smaller than $|D_R|$, the rate of exchange of energy is dominated by the interaction with the eddies which have the longest convective time-scale. Note, however, that the analysis presented here does not apply when $t_{\rm conv} \sim t_{\rm osc}$, so there is still a possibility that resonant interaction between tides and convection plays a significant role.

In the case of fast tides, since the tidal oscillations are the rapid fluctuations, $D_R > 0$ corresponds to a transfer of energy from the tidal flow to convection, whereas $D_R < 0$ correspond to the tidal flow extracting kinetic energy from convection. If the tide is approximated by its equilibrium value, $\langle u'_i u'_j \rangle > 0$ and therefore tidal dissipation requires the gradient of the convective velocity to couple to this Reynolds stress in a particular way. As in the case of slow tides, it is difficult to envision how the tidal oscillation could induce such a coupling.

However, like for the case of slow tides, Terquem & Martin (2021) showed that assuming $D_R > 0$ in the regime $t_{\rm conv} \gg t_{\rm osc}$ yields dissipation rates which account for the circularization of solar-type binaries, which has been a longstanding theoretical puzzle.

## 5 PHASE LAG AND TIDAL DISSIPATION FACTOR

The tidal dissipation factor is only relevant when the tide is locally dissipated by interaction with convection, i.e. when $D_R > 0$ for fast tides or $D_{R,\,\rm slow} < 0$ for slow tides. Therefore, in this section, we assume that the tide is indeed locally dissipated and derive an expression for the phase lag and tidal dissipation factor using the formalism presented above.

### 5.1 Phase lag

As already mentioned, when there is dissipation, the Lagrangian displacement $\boldsymbol{\xi}$ lags behind the tidal force $\boldsymbol{f}_t$ by a phase $\delta$. Therefore, if the tidal potential is proportional to $\cos(2\varphi - \omega_{\rm osc} t)$ in the rotating frame, then $u'_r$ and $u'_\theta$ are proportional to $\sin(2\varphi - \omega_{\rm osc} t - \delta)$ while $u'_\varphi$ is proportional to $\cos(2\varphi - \omega_{\rm osc} t - \delta)$. (We are assuming here that the phase shift is the same for all the components of $\boldsymbol{\xi}$, which may not actually be the case; e.g. Bunting, Papaloizou & Terquem 2019). This yields $\langle \boldsymbol{f}_t \cdot \boldsymbol{u}' \rangle \sim f_t(r) u'(r) \sin\delta$ where $f_t(r)$ and $u'(r)$ denote positive characteristic values of $f_t$ and $u'$ at $r$ and where we assume averages over $\theta$. From equation (29), we then have $\delta \sim D_R / [f_t(r) u'(r)]$, where we have used $\sin\delta \simeq \delta$ as dissipation is weak. Equation (37) gives $D_R \sim u'^2(r)/t_{\rm conv}(r)$, where again we assume averages over $\theta$. This then yields $\delta \sim u'(r)/[f_t(r) t_{\rm conv}(r)]$.

For the equilibrium tide, the characteristic value of the tidal displacement is $\xi(r) \sim \Psi_t(r)/g(r)$ (Ogilvie 2014). Since $u'(r) = \omega_{\rm osc} \xi(r)$, we then get

$$\delta(r) \sim \frac{r \omega_{\rm osc}}{g(r) t_{\rm conv}(r)}. \tag{38}$$

Note that, here, we rely on first-order perturbation theory to compute the phase shift, i.e. we use the velocity calculated while ignoring dissipation to derive the phase shift that results from dissipation. This approach, which was also used in Terquem et al. (1998), is valid because the energy dissipated during a tidal cycle is small compared to the energy contained in the tides.

If the mass $M$ of the star or planet is centrally condensed, then $g(r) \simeq GM/r^2$. As already mentioned above, this is a very good approximation in the convective envelope of the Sun. It is also a reasonably good approximation in the outer parts of the envelope of Jupiter and Saturn, where the tides are significant. Therefore, the phase shift can be approximated as

$$\delta(r) \sim \frac{r^3}{GM} \frac{\omega_{\rm osc}}{t_{\rm conv}(r)}. \tag{39}$$

This is exactly the same expression as that obtained by Darwin (1879) for a viscous sphere, except that in Darwin's expression $r$ is the radius of the sphere and $t_{\rm conv}$ is a coefficient which is assumed to be uniform. Zahn (1977) later used Darwin's formula by identifying this coefficient with what he called the 'friction time', which is the time it takes for convection, assumed to act as a turbulent viscosity, to transport energy throughout the convective envelope of the star.

In Section 3.4, we calculated that the ratio of the work done by the tidal force on the convective flow to that done on the oscillation is $\eta = \epsilon V/(g t_{\rm conv} \delta)$, where $\epsilon$ is given by equation (20). Using equation (38), this yields $\eta \sim \epsilon \lambda_{\rm conv} t_{\rm osc}/(r t_{\rm conv}) \ll 1$, which confirms that most of the work is done on the oscillation.

If the amplitude of inertial waves dominates over that of the equilibrium tide, then $u'(r) \sim 2\Omega \xi(r)$, where $\xi$ is the displacement corresponding to the equilibrium tide (see Section 3.8). Therefore, equations (38) and (39) still apply but with $\omega_{\rm osc}$ being replaced by $2\Omega$, with the caveat that possible strong latitudinal dependence of inertial waves are not captured by the averaging over $\theta$.

The time lag $\Delta t$ is defined through $\delta(r) = \omega_{\rm osc} \Delta t(r)$, and therefore equation (38) yields a time lag independent of frequency for the equilibrium tide.

The expression above has been derived assuming fast tides. For slow tides, $D_R$ is $(\lambda_{\rm conv}/\lambda_{\rm osc})^2$ smaller, so that the right-hand side of equations (38) and (39) has to be multiplied by this factor.

In previous studies, it has been assumed that tidal dissipation in a giant planet or in a star could be quantified by a single uniform phase lag. However, equation (38) shows that $\delta$ *is strongly dependent on $r$*, increasing sharply towards the surface. The phase lag is uniform when the response of the body is viscoelastic, but not in the presence of a fluid.

### 5.2 Phase lags associated with different satellites

Let us consider the case where there are two moons, contributing tidal forces per unit mass $\boldsymbol{f}_{t,1}$ and $\boldsymbol{f}_{t,2}$ with frequencies in the rotating frame $\omega_{\text{osc},1}$ and $\omega_{\text{osc},2}$, respectively. Then $\boldsymbol{u}' = \boldsymbol{u}'_1 + \boldsymbol{u}'_2$, where $\boldsymbol{u}'_i$ (with $i = 1, 2$) oscillates with frequency $\omega_{\text{osc},i}$. Averaging over a time long compared to both $2\pi/\omega_{\text{osc},1}$ and $2\pi/\omega_{\text{osc},2}$ removes the terms involving a product of perturbations with different frequencies (even if the frequencies are commensurate). Equation (29) then becomes

$$\langle \boldsymbol{f}_{t,1} \cdot \boldsymbol{u}'_1 \rangle + \langle \boldsymbol{f}_{t,2} \cdot \boldsymbol{u}'_2 \rangle = D_{R,1} + D_{R,2}, \tag{40}$$

with $D_{R,1} = \langle u'_{1,i} u'_{1,j} \rangle (\partial V_i / \partial x_j)$ and similarly for $D_{R,2}$.

Because, on average, $\boldsymbol{f}_{t,1}$ does not work on $\boldsymbol{u}'_2$, and $\boldsymbol{f}_{t,2}$ does not work on $\boldsymbol{u}'_1$, the dissipation rate $D_{R,i}$ can only be due to the work done by $\boldsymbol{f}_{t,i}$ on $\boldsymbol{u}'_i$. The equation above then implies that $\langle \boldsymbol{f}_{t,i} \cdot \boldsymbol{u}'_i \rangle = D_{R,i}$ for both $i = 1$ and $i = 2$, so that equation (38) is satisfied for each of the moons.

### 5.3 Harmonic oscillator and tidal dissipation factor

We now focus on the equilibrium tide. Ever since Goldreich (1963), Kaula (1964), and MacDonald (1964), the tidal dissipation factor $Q$ has been extensively used to quantify the amount of energy dissipation in moons, planets and stars.

In Appendix B, we give a brief review of the calculation of the phase lag $\delta$ and $Q$ factor for a driven and damped harmonic oscillator. Comparing the expressions obtained for $\sin \delta$ in both the case of the equilibrium tide (equation 38) and the harmonic oscillator (equation B4), we see that we can model the equilibrium tide as a harmonic oscillator with natural frequency $\omega_0 = (g/r)^{1/2} \simeq (GM/r^3)^{1/2}$, driving frequency $\omega = \omega_{\text{osc}}$ and damping coefficient $\gamma = 1/t_{\text{conv}}$, providing $\omega_0^2 \gg \omega_{\text{osc}}^2 \gg \gamma \omega_{\text{osc}}$. For both Saturn and Jupiter interacting with their closest moons, these inequalities are well satisfied, although in the outer parts of the envelope $\omega_0^2$ is only about 5 times $\omega_{\text{osc}}^2$. For Saturn interacting with Rhea and Titan, this approximation is only marginally satisfied at the surface of the envelope, where $\omega_0^2/\omega_{\text{osc}}^2 \sim 2$.

We can therefore approximate the equilibrium tide at some location $r$ as a one dimensional harmonic oscillator which equation of motion is

$$\frac{\mathrm{d}^2 \xi}{\mathrm{d}t^2} + \omega_0^2 \xi = -\gamma u' + f_t \cos(\omega_{\text{osc}} t), \tag{41}$$

where $\xi$ and $u'$ are characteristic values at $r$ of (any component of) the tidal displacement and velocity, respectively, and $f_t$ is a characteristic value at $r$ of the amplitude of the tidal force per unit mass. In this equation, $-\omega_0^2 \xi$ and $-\gamma u'$ are the restoring and friction forces per unit mass, respectively. The restoring force is the gravitational force since $\omega_0^2 \xi \equiv g\xi/r = gu'/(\omega_{\text{osc}} r) \sim \rho'g/\rho_0$, where the mass conservation (19) has been used. For a harmonic oscillator, the rate of work done by the friction force, $\gamma u'^2$, is equal to the rate of work done on average over a period by the driving force. As $\gamma u'^2 = D_R$, identifying $\gamma u'$ with the friction force is consistent with equation (29).

The equilibrium tide assumes that the tidal displacement adjusts itself so that the flow is always at equilibrium in the perturbing potential. This is only satisfied if $\left| \mathrm{d}^2 \xi / \mathrm{d}t^2 \right| \ll \left| \omega_0^2 \xi \right|$, i.e. $\omega_{\text{osc}}^2 \ll \omega_0^2$, which is consistent with the condition required for $\sin \delta$ to be approximated by equation (B5). In that case, the energy of the oscillator is dominated by the potential energy, which is $e'_p(r) = \rho_0 \omega_0^2 \xi^2/2 \sim \rho' g \xi \sim \rho' \Psi_t$ per unit volume, so that the peak energy $E^\star$ in the expression (B8) of the $Q$ factor is the peak potential energy, not the peak kinetic energy.

It is clear from the discussion above that the equilibrium tide can only be described *locally* as a harmonic oscillator.

Fig. 1 shows the time lag $\Delta t(r) = \delta(r)/\omega_{\text{osc}}$ for both Jupiter and Saturn and $Q(r) \equiv 1/\sin \delta(r)$ for Jupiter interacting with Io and for Saturn interacting with Enceladus as a function of position in the planet. The models for Jupiter and Saturn have been provided by I. Baraffe (Baraffe, Chabrier & Barman 2008). The model for Saturn has a ratio of the mixing length to pressure scale height $\alpha = 0.5$, instead of the value of 2 used in stars, as it has been argued that this may be better suited for planetary interiors (see Terquem 2021 for a more detailed discussion about these models). As can be seen on this figure, $\delta$, and therefore $Q$, vary by orders of magnitude throughout the envelope.

### 5.4 Comparison with observations for Jupiter and Saturn

Observational constraints on tidal dissipation in a planet are obtained by calculating the tidal deformation of the planet due to a moon, and the gravitational potential $V_{\text{ext}}$ that this deformation produces at the position of the moon. It is assumed that, because of tidal dissipation in the planet, the tidal bulge (equilibrium tide) lags behind the line joining the centres of the planet and moon, which corresponds to a phase lag $\delta$. Therefore, $V_{\text{ext}}$ results in the planet exerting a torque onto the moon. The component $\Gamma_z$ of this torque along the direction of the orbital angular momentum yields a secular acceleration of the moon. Since $\Gamma_z \propto k_2 \sin \delta$, where $k_2$ is the Love number of the planet, comparing the solutions of the equations of orbital evolution with the observations enables constraints to be put on the phase shift ($k_2$ being independently constrained). The tidal dissipation factor $Q$ is calculated through $Q = 1/\tan \delta$. More details can be found in Goldreich & Soter (1966), Mignard (1980), Murray & Dermott (1999), and Lainey, Dehant & Pätzold (2007).

When more than one moon is present, each moon $i$ is assumed to be associated with a distinct phase shift $\delta_i$. This corresponds to a potential $V_{\text{ext},i}$ produced by the planet at the position of all the moons, but this potential is only associated with a torque on moon $i$ when an average is taken over a time-scale long compared to all the tidal periods. Constraints are derived by taking into account the interactions between all the objects in the system, and the secular acceleration of the different moons has contribution from dissipation in the planet but also in the moons themselves.

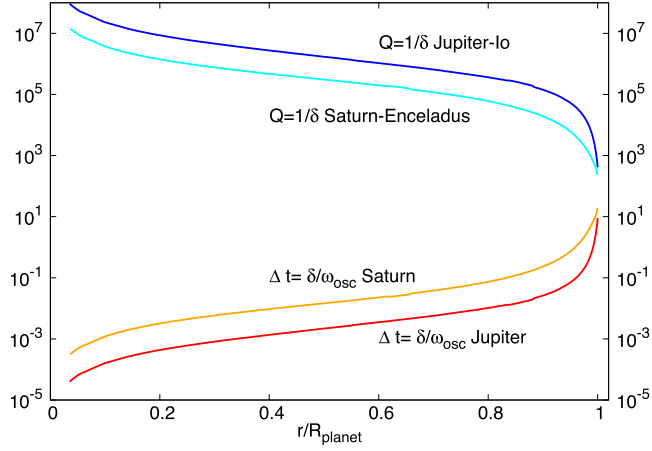

**Figure 1.** Time lag $\Delta t = \delta/\omega_{\rm osc}$ for Jupiter (red curve) and Saturn (orange curve) and tidal dissipation factor $Q \equiv 1/\sin \delta$ for the tides raised on Jupiter by Io (blue curve) and the tides raised on Saturn by Enceladus (cyan curve), in logarithmic scale, *versus* $r/R_{\rm planet}$, where $R_{\rm planet}$ is either the radius of Jupiter or that of Saturn. $\Delta t$ is independent of $\omega_{\rm osc}$ whereas $Q \propto 1/\omega_{\rm osc}$. The phase lag $\delta$ varies by orders of magnitude through the envelope of the planets. The observations give $Q = 3.56 \times 10^4$ for Jupiter interacting with Io, and $Q = 2.45 \times 10^3$ for Saturn interacting with Enceladus.

Although these studies were initially developed for rocky planets, for which the response is viscoelastic and therefore well described by a uniform phase shift, they have also been used to quantify tidal dissipation in giant planets (Lainey et al. 2009, 2012; Jacobson 2022).

As shown above, $\delta$ for a gaseous planet is not uniform, and varies by orders of magnitude throughout the planet. In order to relate the analysis done in this paper to observational constraints, we now show that we can define an average phase shift for the planet. The torque $\Gamma_z$ exerted by the planet on the moon is equal and opposite to that exerted by the moon on the planet. Therefore

$$\Gamma_z = \int_{\mathcal{V}} \rho' \frac{\partial \Psi_t}{\partial \varphi} {\rm d}v, \qquad (42)$$

where $\mathcal{V}$ is the volume of the envelope of the planet.

For a circular orbit, the tidal potential is given by $\Psi_t(r, \theta, \varphi, t) = 3 f r^2 \sin^2 \theta \cos(2\varphi - \omega_{\rm osc} t)$ with $f = -GM_p/(4\, a^3)$, where $M_p$ is the mass of the companion which excites the tides and $a$ is the binary separation. We have shown in Section 3.6 that $\langle \rho' \boldsymbol{g} \cdot \boldsymbol{u}' \rangle$ was negligible. This implies that the phase shift between $\rho'$ and $\xi_r$ is very small compared to the phase shift $\delta$ between $\xi_r$ and the tidal potential (otherwise $\langle \rho' \boldsymbol{g} \cdot \boldsymbol{u}' \rangle$ would be comparable to $\rho_0 \langle \boldsymbol{f}_t \cdot \boldsymbol{u}' \rangle$, and therefore to $\rho_0 D_R$). Therefore, $\rho'$ lags behind $\Psi_t$ by $\delta$ and we can write $\rho'(r, \theta, \varphi, t) = \rho'(r) h(\theta) \cos(2\varphi - \omega_{\rm osc} t - \delta)$, with $h$ a function of $\theta$. We have assumed here that the variables are separable, which is not the case when there is rotation. However, for uniform rotation (applicable for most of the interior of Jupiter and Saturn), the tidal displacement is well approximated by that corresponding to a non-rotating body (Ioannou & Lindzen 1993). Using these expressions of $\Psi_t$ and $\rho'$ in equation (42) then yields

$$\Gamma_z = -6\pi f \int_0^{\pi} \sin^3 \theta h(\theta) {\rm d}\theta \int_{R_i}^{R_p} \rho'(r) r^4 \sin \delta(r) {\rm d}r, \qquad (43)$$

where $R_i$ and $R_p$ are the inner and outer radii of the envelope, respectively. We can write $\Gamma_z$ as being proportional to an average $\sin \delta$ for the planet by defining this average as

$$\overline{\sin \delta} = \frac{\int_{R_i}^{R_p} \rho'(r) r^4 \sin \delta(r) {\rm d}r}{\int_{R_i}^{R_p} \rho'(r) r^4 {\rm d}r}. \qquad (44)$$

This yields an average time lag $\overline{\Delta t} = \overline{\delta}/\omega_{\rm osc}$ where $\overline{\delta} \simeq \overline{\sin \delta}$.

Since $\sin \delta(r) \sim D_R / [f_t(r) u'(r)] \sim r D_R / [\Psi_t(r) u'(r)]$ and $\rho'(r) \sim \rho_0 u'(r)/(r\omega_{\rm osc})$, equation (44) can also be written as

$$\overline{\sin \delta} \sim \frac{\int_{R_i}^{R_p} \rho_0 D_R r^2 {\rm d}r}{\omega_{\rm osc} \int_{R_i}^{R_p} \rho' \Psi_t(r) r^2 {\rm d}r}. \qquad (45)$$

The numerator is the total energy dissipated per unit time, whereas the integral in the denominator is the total potential energy $E'_p$ in the tide. Therefore, we have $\overline{\sin \delta} \sim 1/Q$ with:

$$Q = \frac{2\pi E'_p}{\Delta E}, \qquad (46)$$

where $\Delta E$ is the total energy dissipated during one oscillation period.

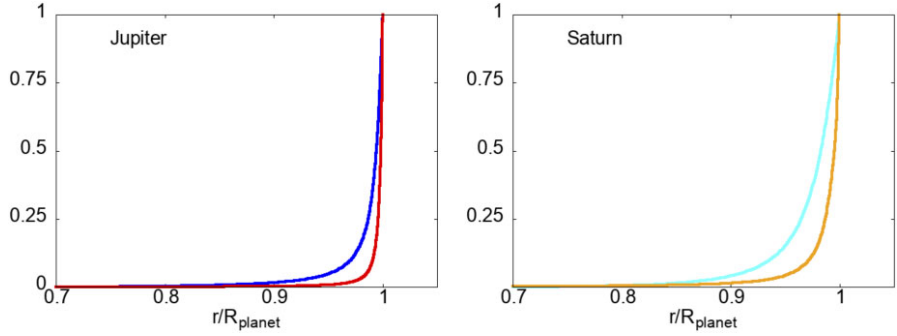

**Figure 2.** Dissipation of the tides raised in Jupiter (left-hand panel) and Saturn (right-hand panel) by Io and Enceladus, respectively. The blue and cyan curves show $\int_{R_i}^{r} \rho'(r) r^4 \sin\delta(r) \, dr$ normalized to unity *versus* $r/R_{\text{planet}}$. The parts of the envelope which contribute most to $\overline{\sin\delta}$ are above the value of $r$ at which this quantity becomes non negligible. Therefore, the curves show that most of the dissipation occurs in the ∼15 outer per cent of the envelopes. The red and orange curves show $1/t_{\text{conv}}$ normalized to unity. This is a measure of the Brunt–Väisälä frequency, and therefore of non-adiabaticity.

Terquem (2021) incorrectly took twice the kinetic energy instead of the potential energy to compute $Q$, arguing equipartition between kinetic and potential energy. Although those energies are comparable in the outer parts of Jupiter's envelope, they are however not the same because $\omega_0$ is a few times $\omega_{\text{osc}}$. This also led an incorrect dependence on the tidal frequency.

To compute the average of $\sin\delta$, equation (44) shows that it has to be weighed by $\rho'$, and not by $\rho_0$. This is because $\sin\delta$, which does not depend on the perturbation, is only relevant in the parts of the flow where the tidal perturbation is significant. The exact form of $\rho'(r)$ is therefore not important when computing $\overline{\sin\delta}$, as long as it tracks the tidal perturbation. To get numerical values of $\overline{\sin\delta}$, we then use $\rho'$ corresponding to the standard equilibrium tide, which is incompressible, i.e. $\rho'(r) = -(d\rho_0/dr)\xi_r(r)$, with $\xi_r(r) = -3fr^2/g(r)$, even though in the regime $t_{\text{osc}} < t_{\text{conv}}$ it is not the correct form of the equilibrium tide (Terquem et al. 1998; Goodman & Dickson 1998). We have checked that using $\xi_r$ to weigh $\sin\delta$ in equation (44), instead of $\rho'$, does not make a difference, which proves that using an approximate form of $\rho'$ is sufficient. To calculate $Q$ directly from $E_p'$ though, as given by equation (46), the correct form of $\rho'$ has to be used. We have indeed checked that the standard equilibrium tide gives a value of $Q$ from equation (46) which is about an order or magnitude larger than $1/\overline{\sin\delta}$ calculated from equation (44).

We now evaluate $\overline{\delta}$ for Jupiter and Saturn.

For the tides raised in Jupiter by any of its moons, equation (44) gives $\overline{\Delta t} = 0.3$ s and, in the case of Io, for which $\omega_{\text{osc}} = 2.7 \times 10^{-4}$ s$^{-1}$, $1/\overline{\delta} = 1.3 \times 10^4$. This is close to the value of $3.56 \times 10^4$ derived by Lainey et al. (2009).

For the tides raised in Saturn by any of its moons, we obtain $\overline{\Delta t} = 1.0$ s and $1/\overline{\delta} = 6 \times 10^3$ for Mimas (for which $\omega_{\text{osc}} = 1.7 \times 10^{-4}$ s$^{-1}$), $1/\overline{\delta} = 4 \times 10^3$ for Enceladus (for which $\omega_{\text{osc}} = 2.2 \times 10^{-4}$ s$^{-1}$), $1/\overline{\delta} = 3 \times 10^3$ for Rhea (for which $\omega_{\text{osc}} = 3.0 \times 10^{-4}$ s$^{-1}$), and $1/\overline{\delta} = 3 \times 10^3$ for Titan (for which $\omega_{\text{osc}} \simeq 2\Omega = 3.2 \times 10^{-4}$ s$^{-1}$).

For Mimas, Enceladus, Tethys, and Dione interacting with Saturn, Lainey et al. (2020), and Jacobson (2022) derived a value of $\Delta t$ roughly between 0.3 and 3.7 from observations, which is in reasonable agreement with our value for $\overline{\Delta t}$. For Titan, Jacobson (2022) also has a value within that range, whereas Lainey et al. (2020) derived a time lag 10 times larger. For Rhea, both studies report values of $\Delta t$ close to 10.

For both Jupiter and Saturn, Fig. 2 shows the regions which contribute most to $\overline{\sin\delta}$, and therefore to tidal dissipation.

In principle, we could also calculate directly the orbital evolution time-scale $t_a = a/(da/dt) = |E_{\text{orb}}|/(dE/dt)$, where $E_{\text{orb}} = -GMM_p/(2a)$ is the orbital energy (with $M$ being the planet's mass and $M_p$ the satellite's mass) and $dE/dt$ is the energy dissipated per unit time. Using $dE/dt = \int \rho_0 \left(u_r'^2/t_{\text{conv}}\right) dv$, where the integral is over the volume of the convective envelope, we obtain $t_a^{-1} = 0.3 \times 10^{-10}$ yr$^{-1}$ for Io. However, this cannot be compared meaningfully to the observations, which give $t_a^{-1} = 0.09 \times 10^{-10}$ yr$^{-1}$ (Lainey et al. 2009), because our calculation does not include the contribution from the dissipation of tidal energy in Io itself, nor the effect of Europa and Ganymede, which is important because of the Laplace resonance the satellites are in. The fact that Io is moving towards Jupiter, instead of away from it as would be the case if only tidal dissipation in the planet were important, shows that the motion of Io is dominated by these other contributions.

## 6 SUMMARY AND DISCUSSION

### 6.1 Summary

The work presented in this paper shows that the energy of a tidal oscillation in a convective flow can only be exchanged with the convective flow by changing the kinetic energy of this flow, not its internal nor potential energy. The analysis has been done for $t_{\text{osc}} \ll t_{\text{conv}}$, and in this case the rate $D_R$ of energy exchange couples the Reynolds stress associated with the oscillation to the gradient of the convective velocity. This result is valid even when the flows are compressible and in the presence of uniform rotation, and applies whether the oscillation is the equilibrium tide or a superposition of the equilibrium tide and a propagating inertial wave. If the oscillation is a $p$ mode, the rate at which the kinetic energy of the oscillation is exchanged with the kinetic energy of the convective flow is still given by $D_R$. However, in that case, and because of compressibility, there is also an exchange between the kinetic energy of the oscillation and the potential and internal energy of convection.

The analysis would still apply when $t_{osc} \gg t_{conv}$, and the rate of energy exchange per unit mass would still be $D_R$, but with this term now coupling the Reynolds stress associated with the convective velocities to the gradient of the velocity of the oscillation.

In the case $t_{osc} \ll t_{conv}$, $|D_R| \sim u'^2 V/\lambda_{conv}$, where $u'$ and $V$ are the velocities of the oscillation and convection, respectively. In the case $t_{osc} \gg t_{conv}$, and assuming that mixing length theory applies in this regime, $|D_R|$ has the same form but is $(\lambda_{conv}/\lambda_{osc})^2$ smaller. Therefore, not only is the energy exchange not suppressed for fast tides, contrary to what has been assumed in previous studies, it is actually much larger than for slow tides! Local dissipation of the oscillation requires $D_R > 0$ when $t_{osc} \ll t_{conv}$ and $D_R < 0$ when $t_{osc} \gg t_{conv}$. This means that whichever flow is varying faster has to transport the momentum associated with the slowly varying flow from regions where it is high to regions where it is lower. It is not clear how, or even if, that happens.

Focusing on tidal oscillations, and assuming local dissipation of the tides, we have calculated the phase lag $\delta(r)$ between the oscillation and the tidal potential. We have shown that this is simply given by $r\omega_{osc}/(gt_{conv})$, where the gravitational acceleration $g$ and $t_{conv}$ have to be evaluated locally. The equilibrium tide can be described locally as a harmonic oscillator with natural frequency $(g/r)^{1/2}$ and subject to a damping force $-u'/t_{conv}$.

Although $\delta(r)$ varies by orders of magnitude through the convective envelope of a planet, it is possible to define an average phase shift $\bar{\delta}$ which can be compared to the phase shift derived from observations. For the equilibrium tide, we have found that $1/\bar{\delta}$ is equal to the standard tidal dissipation factor $Q = 2\pi E'_p/\Delta E$. As $\bar{\delta} \propto \omega_{osc}$, the time lag associated with this phase shift does not depend on frequency (i.e. it is uniquely defined for a planet, independently of the moon that raises the tides), and $Q \propto 1/\omega_{osc}$.

### 6.2 Discussion

It has been proposed that the dissipation of inertial waves could explain the circularization of solar type binaries (Barker 2020, 2022). In these studies, no specific dissipation mechanism is being proposed, it is just assumed that dissipation takes place. As the analysis we have presented above applies to propagating inertial waves as well as to the equilibrium tide, it shows that inertial waves cannot in general be an alternative to the equilibrium tide to explain dissipation: if they are dissipated, the equilibrium tide is dissipated as well.

Of course, inertial waves could in principle provide more dissipation. However, this is not borne out by the results of Barker (2022) for solar-type stars. Their fig. 3 shows that solar-type stars can only reach the circularization periods observed for 10 Gyr clusters if they circularize up to about 9 d on the PMS, which is significantly above the value of 7 d derived from observations.

Therefore, circularization on the PMS is clearly overestimated in this calculation (this is achieved by starting the tidal interaction when the stars are only 0.15 Myr). If circularization is actually only achieved up to orbital periods of 7 d on the PMS, as suggested by observations, significant tidal dissipation is needed after the MS, when the star rotates much more slowly. For the Sun in its current state, inertial waves would only be tidally excited in binaries with orbital periods larger than 13 d. They could therefore not explain the increase of the circularization period from 7 d on the PMS to 10–12 d at the beginning of the RGB. Finally, we note that, in these studies, the energy dissipated by inertial waves is calculated using an average over all frequencies. This formalism was initially proposed by Ogilvie (2013) to calculate the dissipation of energy when the forcing is impulsive. It is appropriate if tides are raised during a brief encounter or in very eccentric orbits, but not in a circular binary when only one dominant frequency contributes to the tides. The reason invoked by Barker (2020) for using this averaging is that the dissipation of energy of inertial waves varies by orders of magnitude depending on the frequency. However, this is by no means a justification for using an averaging over all possible frequencies.

If $D_R$ has the sign required for local dissipation of the oscillation to occur, then dissipation of the equilibrium tide alone explains the circularization periods of solar-type stars derived from observations, and the interaction does not need to be started before about 0.4 Myr for circularization up to periods of 7 d to be achieved on the PMS (Terquem & Martin 2021). For Jupiter and Saturn, the results presented in this paper show that the phase shift due to tidal dissipation of the equilibrium tide is also consistent with observations, except for the tides raised by Rhea in Saturn, and maybe also for the tides raised by Titan in this planet. However, for these moons, the equilibrium tide approximation may not apply in the outer parts of the envelope, which contribute most to dissipation, as the natural frequency $\omega_0$ is comparable to the oscillation frequency $\omega_{osc}$ there, as pointed in Section 5.3. Resonance locking with inertial waves has been proposed as a possible mechanism driving the evolution of the moons of Saturn (Fuller, Luan & Quataert 2016; Lainey et al. 2020), and it explains the phase shift of Rhea and Titan. But again, this can only occur if these inertial waves are dissipated by interaction with convection. Resonance locking also requires the tidal oscillation to resonate with a free inertial wave in the planet. Whether such free modes can be maintained is still an open question.

In this context, it is important to note that mode–mode coupling for inertial waves, whether it is a tidally driven oscillation resonating with a free mode, or free modes parametrically interacting with each other, cannot be modelled in the presence of a turbulent viscosity arising from convection. As demonstrated in this paper, convection does not act as a turbulent viscosity. Instead, damping of an inertial wave which interacts with a convective flow is itself a result of mode–mode coupling between the inertial wave and the unstable gravity waves which characterize convection.

The very important question that remains to be answered is whether $D_R$ has the sign needed for the tidal oscillation to be damped. It has always been assumed to be the case for slow tides, when mixing length theory is used, and it is indeed what numerical simulations show (Ogilvie & Lesur 2012; Duguid, Barker & Jones 2020; Vidal & Barker 2020). For fast tides, there is some suggestion in the simulations performed by Barker & Astoul (2021) that $D_R$ integrated over the flow domain is positive, which corresponds to tidal dissipation. The total rate of energy dissipation is found to be significantly smaller than what we obtain by assuming local dissipation, but the results may be affected by the use of rigid boundary conditions, as discussed in Section 2.2.

Further work, and in particular numerical simulations, are of course needed to investigate the dissipation of tidal oscillations when $t_{\rm osc} \ll t_{\rm conv}$.

As compressiblility has been taken into account, the formalism presented in this paper applies to *p* modes. The oscillation period of *p* modes is much smaller than the convective time-scale of the slowest eddies in a large part of the convective zone, and therefore mixing length theory does not apply to describe the interaction of the modes with these eddies. It has been proposed that the damping of *p* modes is dominated by resonant interactions with convection, i.e. by interactions with eddies which have a convective time-scale comparable to that of the oscillation, and this interaction has been studied using the mixing length approximation (Goldreich & Keeley 1977). However, the analysis presented here shows that, if this approximation applies, it is only in the regime $t_{\rm conv} \ll t_{\rm osc}$. Numerical simulations actually confirm that mixing length theory is not a good approximation to model the damping of *p* modes (Basu 2016). Resonant interaction is not captured by the analysis presented in this paper, which relies on a separation of time-scales. If important, resonant interaction needs to be investigated using a different approach. It would still be interesting to study the interaction of *p* modes with the slowest eddies, to obtain some estimate of the energy damping rate to which they contribute. Existing theories are indeed not fully successful at reproducing the mode linewidths (Houdek & Dupret 2015), and it has been argued that a novel approach is needed (Belkacem et al. 2019).

## ACKNOWLEDGEMENTS

I thank Steven Balbus for his support, his patience in answering questions and very insightful comments that have been invaluable in shaping the work presented in this paper. I am grateful to John Papaloizou for commenting on various parts of this work, and sharing his vast knowledge of the topic. I also thank David Marshall and Michael McIntyre for giving me some interesting perspectives from oceanography and atmospheric sciences, and Isabelle Baraffe, Sacha Brun, Antoine Strugarek, and Dimitar Vlaykov for enlightening discussions about convection in the Sun. I have also benefited greatly from very stimulating discussions with all the participants of the *Tidal Evolution Research Review for Astrophysics (TERRA)* workshop, which was held in Leiden in January 2023, and particularly with Dong Lai, Valery Lainey, Gordon Oglvie and Jack Wisdom. Finally, I thank Scott Tremaine and Kévin Belkacem for comments on an earlier draft of this paper, which led to some important corrections, and an anonymous referee whose very thoughtful reviews resulted in very significant improvements.

## DATA AVAILABILITY

No new data were generated or analysed in support of this research.

  

## APPENDIX A: TIME AVERAGING AND THE ROLE OF VELOCITY CORRELATIONS

This appendix discusses the effect of time averaging on the rate of energy exchange between the fluctuations and the mean flow.

Here, we use a Reynolds decomposition $\boldsymbol{u} = \boldsymbol{V} + \boldsymbol{u}'$ where the velocity $\boldsymbol{u}'$ varies on a time-scale $t_1$ and the velocity $\boldsymbol{V}$ varies on a time-scale $t_2 \gg t_1$. If $t_1 = t_{\rm osc}$ and $t_2 = t_{\rm conv}$, then we are in the regime of fast tides, and $\boldsymbol{u}'$ is the tidal velocity (fluctuations) whereas $\boldsymbol{V}$ is the convective velocity (mean flow). However, if $t_1 = t_{\rm conv}$ and $t_2 = t_{\rm osc}$, then we are in the regime of slow tides, and $\boldsymbol{u}'$ is the convective velocity (fluctuations) whereas $\boldsymbol{V}$ is the tidal velocity (mean flow). Although the paper focusses on fast tides, the discussion in this appendix is valid for both cases. When dealing with slow tides, we assume that there exists a time $\tau$ such that $t_1 = t_{\rm conv} \ll \tau \ll t_2 = t_{\rm osc}$. When dealing with fast tides, we take $\tau = t_1 = t_{\rm osc}$. In both cases, we also define a time $\tau' \gg t_2$. We note $<\ldots>$ a time average over $\tau$ and $\langle \ldots \rangle_{\tau'}$ a time average over $\tau'$. In this appendix, we also assume that both the mean flow and the fluctuations are incompressible, as this makes the discussion simpler and it does not affect the argument presented here.

To derive an equation for the kinetic energy of the fluctuations, we dot Navier–Stokes equation with $\boldsymbol{u}'$, which brings in the term $\boldsymbol{u}' \cdot (\boldsymbol{u} \cdot \nabla) \boldsymbol{u}$. Using the Reynolds decomposition and $\nabla \cdot \boldsymbol{u}' = \nabla \cdot \boldsymbol{V} = 0$, this term can be written as the sum of a divergence and $d_R$, where we define

$$d_R = u'_i u'_j \frac{\partial V_i}{\partial x_j} - V_i V_j \frac{\partial u'_i}{\partial x_j}. \tag{A1}$$

The divergence term does not exchange energy between the fluctuations and the mean flow, it only transports energy through the flow, so that we ignore it. Therefore, the term $(\boldsymbol{u} \cdot \nabla) \boldsymbol{u}$ in Navier–Stokes equation only contributes $d_R$ to the kinetic energy of the fluctuations. If we now average over the time $\tau$, during which $V_i$ and $V_j$ stay almost constant, we obtain

$$\langle d_R \rangle = D_R = \langle u'_i u'_j \rangle \frac{\partial V_i}{\partial x_j}. \tag{A2}$$

Therefore, $D_R$ is the only term contributed by $(\boldsymbol{u} \cdot \nabla) \boldsymbol{u}$ to the mean kinetic energy of the fluctuations.

Now, as we are usually interested in the energy exchange between the fluctuations and the mean flow over a long time-scale, $D_R$ has to be averaged over $\tau'$. Naively, we could expect $\langle D_R \rangle_{\tau'} = 0$, because $\langle \partial V_i / \partial x_j \rangle_{\tau'} = 0$. However, in a standard shear flow, where the turbulence is due to instabilities in the flow itself, the energy of the turbulent eddies has to come from the free energy present in the shear. It is therefore assumed that $D_R$ has the sign required for energy to be transferred from the mean shear flow to the turbulent fluctuations. In other words, the sign of $\langle u'_i u'_j \rangle$ is correlated with that of $\partial V_i / \partial x_j$, so that $\langle D_R \rangle_{\tau'} \neq 0$.

In the case, where the turbulence is produced by buoyancy, the energy in the turbulent eddies does not depend on a background shear flow. However, mixing length theory assumes that, if a mean shear flow is introduced in the convective flow, energy is transferred from the shear flow to the convective eddies so that here again $\langle D_R \rangle_{\tau'} \neq 0$.

## APPENDIX B: QUALITY FACTOR OF A HARMONIC OSCILLATOR

This appendix reviews the harmonic oscillator and the relation between the quality factor and phase lag.

We consider a harmonic oscillator consisting of a mass $m$ which moves along the $x$-axis and is subject to a restoring force $-kx$, a friction force $-bv$, and a periodic driving force $F = F_0 \cos(\omega t)$, where $k$ and $b$ are positive constants, $F_0$ and $\omega$ are the amplitude and frequency of the forcing, respectively, $x$ is the displacement from equilibrium and $v$ is the velocity. The equation of motion is

$$m \frac{d^2 x}{dt^2} + kx = -bv + F_0 \cos(\omega t), \tag{B1}$$

which general solution is

$$x(t) = A e^{-\gamma t/2} \cos(\omega_1 t + \phi) + X_0 \cos(\omega t - \delta), \tag{B2}$$

where $A$ and $\phi$ are two constants which depend on the initial conditions, $\gamma = b/m$, $\omega_1^2 = \omega_0^2 - \gamma^2/4$ with $\omega_0^2 = k/m$, and

$$X_0 = \frac{F_0/m}{\left[ (\omega_0^2 - \omega^2)^2 + \gamma^2 \omega^2 \right]^{1/2}}, \tag{B3}$$

$$\sin \delta = \frac{\gamma \omega}{\left[ (\omega_0^2 - \omega^2)^2 + \gamma^2 \omega^2 \right]^{1/2}}. \tag{B4}$$

As expected, $\delta > 0$, i.e. the oscillation lags behind the driving force because of dissipation.

If $\omega_0^2 \gg \omega^2 \gg \gamma^2 \omega^2$, then:

$$\sin \delta \simeq \frac{\gamma \omega}{\omega_0^2}, \tag{B5}$$

whereas, if $\omega^2 \gg \omega_0^2 \gg \gamma^2 \omega^2$, then:

$$\sin \delta \simeq \frac{\gamma}{\omega}. \tag{B6}$$

Multiplying equation (B1) by $v$ yields the equation for the conservation of energy

$$\frac{d}{dt} [K(t) + U(t)] = -bv^2 + F_0 v \cos(\omega t), \tag{B7}$$

where $K(t) = mv^2/2$ and $U(t) = kx^2/2$ are the kinetic and potential energy, respectively.

After a time $t \gg \gamma^{-1}$, the first (transient) term in the expression (B2) for $x(t)$ becomes negligible, and $x$ is equal to the steady state solution $X_0 \cos(\omega t - \delta)$. In this regime, the stored kinetic energy is $K(t) = m X_0^2 \omega^2 \sin^2(\omega t - \delta)/2$ and the stored potential energy is $U(t) = m X_0^2 \omega_0^2 \cos^2(\omega t - \delta)/2$. They both remain constant on average over a period, which implies from equation (B7) that the work done by the driving force on average over a period is all done against the friction force. The total energy stored in the oscillator, $E(t) = K(t) + U(t)$, was built to its steady state value during the initial transient phase when only part of the work done by the driving force was acting against the friction force.

The quality factor for a harmonic oscillator is defined as

$$Q = \frac{2\pi E^\star}{\Delta E}, \tag{B8}$$

where $\Delta E$ is the (positive) energy dissipated during one period and $E^\star$ is either the average energy $<E>$ stored in the oscillator during one period (Feynman 1964; Kleppner & Kolenkow 2013), or the peak (maximum) energy stored during one period. In general, the former definition is used for mechanical oscillators whereas the latter is used for electrical oscillators. However, in problems involving tidal dissipation, it is customary to use for $E^\star$ the peak energy.

In steady state, the period is $T = 2\pi/\omega$ and:

$$\Delta E \equiv -\int_0^T \frac{dE}{dt} \, dt = -\frac{2\pi}{\omega} \left\langle \frac{dE}{dt} \right\rangle. \tag{B9}$$

As discussed above, we have $<dE/dt> = -<Fv>$, which yields

$$\Delta E = \pi F_0 X_0 \sin \delta. \tag{B10}$$

The maximum value that $E(t)$ reaches during one period is $E^\star = m X_0^2 \max(\omega^2, \omega_0^2)/2$. Therefore, we obtain

$$Q = \frac{\max(\omega^2, \omega_0^2)}{\left[ (\omega_0^2 - \omega^2)^2 + \gamma^2 \omega^2 \right]^{1/2}} \frac{1}{\sin \delta}. \tag{B11}$$

Note that, using equation (B4), we can also write $Q = \max(\omega^2, \omega_0^2)/(\gamma \omega)$. This is the same expression we obtain by calculating $Q$ directly from its definition and with $dE/dt = -bv^2$.

If $\omega_0^2 \gg \omega^2 \gg \gamma^2 \omega^2$, then:

$$Q \simeq \frac{1}{\sin \delta} \simeq \frac{\omega_0^2}{\gamma \omega}, \tag{B12}$$

whereas, if $\omega^2 \gg \omega_0^2 \gg \gamma^2\omega^2$, then:

$$Q \simeq \frac{1}{\sin \delta} \simeq \frac{\omega}{\gamma}. \tag{B13}$$

When $\omega_0^2 \gg \omega^2 \gg \gamma^2\omega^2$, $E^\star$ is equal to the maximum of the potential energy, which is reached at $t = T/4$ for our choice of initial conditions, and can be calculated while neglecting dissipation. Therefore

$$E^\star \simeq \int_0^{T/4} Fv \, \mathrm{d}t = \frac{F_0 X_0}{2} \left( \cos \delta - \frac{\pi}{2} \sin \delta \right). \tag{B14}$$

Using $|\sin \delta| \ll |\cos \delta|$ together with equations (B8) and (B10) then yields

$$Q \simeq \frac{1}{\tan \delta}. \tag{B15}$$

This paper has been typeset from a TeX/LaTeX file prepared by the author.